\documentclass[preprint,3p,times]{elsarticle}
\journal{Applied Energy}

\def\BibTeX{{\rm B\kern-.05em{\sc i\kern-.025em b}\kern-.08em
    T\kern-.1667em\lower.7ex\hbox{E}\kern-.125emX}}

\usepackage{amsmath, amssymb, amsfonts, mathtools}
\usepackage{graphicx}
\usepackage{textcomp}
\usepackage{booktabs}
\usepackage{tabularx}
\usepackage{array}
\usepackage{float}
\usepackage{url}
\usepackage{hyperref}
\usepackage{xurl}
\hypersetup{breaklinks=true,
  pdftitle={LLMs and Agentic AI Systems for Smart Grids: A Tutorial on Architectures and Applications},
  pdfauthor={Daniela Rojas et al.}}
\usepackage{xcolor}
\usepackage{pifont}
\usepackage{multirow}

\usepackage{tikz}
\usetikzlibrary{positioning, arrows.meta, shapes, shapes.geometric, fit, calc}

\usepackage{tcolorbox}
\tcbuselibrary{skins,breakable}
\usepackage[scaled=0.92]{inconsolata}
\usepackage{microtype}
\usepackage{soul}

\usepackage{silence}
\definecolor{rolebg}{RGB}{226,238,255}
\definecolor{contextbg}{RGB}{226,242,226}
\definecolor{taskbg}{RGB}{255,239,210}
\definecolor{outputbg}{RGB}{240,228,246}
\definecolor{headbg}{RGB}{248,248,248}

\newtcolorbox{promptfigure}[2][]{%
  enhanced,
  sharp corners,
  colback=white,
  colframe=black,
  colbacktitle=headbg,
  coltitle=black,
  boxrule=0.5pt,
  title={#2},
  fonttitle=\bfseries,
  left=0.8mm,right=0.8mm,top=0.4mm,bottom=0.4mm,
  boxsep=0.3mm,
  before skip=2pt,
  after skip=2pt,
  colbacklower=gray!2,
  segmentation style={solid, black, line width=0.35pt},
  before upper={\linespread{0.4}\selectfont},
  before lower={\linespread{0.4}\selectfont},
  #1
}

\newtcbox{\legendchip}[1][]{%
  on line,
  boxrule=0.35pt,
  colframe=black,
  colback=gray!15,
  arc=1pt,
  left=2pt,right=2pt,top=0.5pt,bottom=0.5pt,
  boxsep=0pt,
  fontupper=\scriptsize\bfseries,
  #1
}

\newcommand{\outputline}[1]{{\ttfamily #1}\par}

\definecolor{exonebg}{RGB}{232,241,255}
\definecolor{extwobg}{RGB}{232,247,236}
\definecolor{targetbg}{RGB}{255,243,220}
\definecolor{reasonbg}{RGB}{255,239,210}
\definecolor{retrievedbg}{RGB}{236,232,248}
\definecolor{querybg}{RGB}{233,244,255}

\newcommand{\rolehl}[1]{{\sethlcolor{rolebg}\hl{#1}}}
\newcommand{\contexthl}[1]{{\sethlcolor{contextbg}\hl{#1}}}
\newcommand{\taskhl}[1]{{\sethlcolor{taskbg}\hl{#1}}}
\newcommand{\outputhl}[1]{{\sethlcolor{outputbg}\hl{#1}}}
\newcommand{\exonehl}[1]{{\sethlcolor{exonebg}\hl{#1}}}
\newcommand{\extwohl}[1]{{\sethlcolor{extwobg}\hl{#1}}}
\newcommand{\targethl}[1]{{\sethlcolor{targetbg}\hl{#1}}}
\newcommand{\reasonhl}[1]{{\sethlcolor{reasonbg}\hl{#1}}}
\newcommand{\retrievedhl}[1]{{\sethlcolor{retrievedbg}\hl{#1}}}
\newcommand{\queryhl}[1]{{\sethlcolor{querybg}\hl{#1}}}

\newcommand{\goodmark}{\textcolor{green!50!black}{\ding{51}}}
\newcommand{\badmark}{\textcolor{red!70!black}{\ding{55}}}
\newcommand{\neutralprompt}[1]{{\ttfamily #1\par\vspace{0.2mm}}}

\usepackage{ragged2e}
\newcolumntype{Y}{>{\RaggedRight\arraybackslash}X}
\newcolumntype{P}[1]{>{\RaggedRight\arraybackslash}p{#1}}

\begin{document}

\begin{frontmatter}

\title{LLMs and Agentic AI Systems for Smart Grids: \\ A Tutorial on Architectures and Applications}

\author[ucsd]{Daniela Rojas\corref{cor1}}
\ead{d2rojas@ucsd.edu}
\author[ucsd]{Abdulwahab Albassam}
\author[ucsd]{Aidan G. Leung}
\author[ucsd]{Jett Ngo}
\author[ucsd]{Ryan Luo}
\author[ucsd]{Peter R. Quawas}
\author[ucsd]{Junpyung Kim}
\author[ucsd]{Kangkai Liang}
\author[ucsd]{Mansi Nanavati}
\author[ucsd]{Jonathan Mai}
\author[ucsd]{Meng-Chi Tsai}
\author[ucsd]{Yun-Tong Tsai}
\author[ualberta]{Yize Chen}
\author[ucsd]{Yuanyuan Shi\corref{cor1}}
\ead{yyshi@ucsd.edu}
\affiliation[ucsd]{organization={Department of Electrical and Computer Engineering, University of California San Diego}, city={La Jolla}, state={CA}, postcode={92093}, country={USA}}
\affiliation[ualberta]{organization={Department of Electrical and Computer Engineering, University of Alberta}, city={Edmonton}, state={AB}, postcode={T6G 1H9}, country={Canada}}
\cortext[cor1]{Corresponding authors.}

\begin{abstract}
Large language models (LLMs) and agentic AI systems have evolved from natural language tasks to using external tools to plan, retrieve, and act in technical domains. In smart grids, recent work applies agentic schemes to forecasting, optimization, and control, wrapping trusted solvers behind language interfaces and orchestrating multi-step workflows. The literature lacks a unified approach to designing and evaluating such systems. LLMs can produce numerically plausible yet physically infeasible outputs, evaluation protocols vary across tasks, and the boundary between what the model should and should not compute is implicit. This paper presents a \emph{solver-grounded design} principle: a numerical result is reported only when it originates from a trusted tool and passes explicit verification. We review the building blocks of LLM and agentic AI systems for power systems: prompting strategies and agentic architectures. We instantiate the principle in four case studies: wind power forecasting, EV charging scheduling, power flow analysis, and contingency diagnosis, each comparing an LLM-only baseline against its solver-grounded counterpart on identical data and metrics. EVAgent reproduces the CVXPY optimum while reducing LLM-only unmet energy by 7.5--9.5$\times$, and GridDebugAgent repairs 17/39 contingency cases while reducing total violations by 52.3\%. We propose a four-group evaluation framework spanning task utility, solver-grounded correctness, faithfulness and safe failure, and cost and latency. A consistent division of labor emerges, the agentic system reliably orchestrates, retrieves, and explains, while trusted tools compute and a verification gate decides what is reported.
\end{abstract}


\end{frontmatter}

\section{Introduction}
Adoption of large language models (LLMs) and agentic AI systems has accelerated across research and industry. Google alone reported processing over 3.2 quadrillion tokens per month across its products in mid-2026, a sevenfold year-over-year increase~\cite{pichai2026io}, and Gartner projects that 40\% of enterprise applications will embed task-specific AI agents by the end of 2026, up from less than 5\% a year earlier, with agentic AI accounting for an estimated \$450 billion in enterprise-software revenue by 2035~\cite{gartner2025agentic}. These systems are now widely adopted for software engineering~\cite{chen2021codex}, scientific discovery~\cite{boiko2023autonomous}, robotics~\cite{ichter2023saycan}, and domain-specific decision support~\cite{singhal2023medpalm}. Agentic systems combine LLMs with external tools, memory, retrieval, code execution, and trusted solvers so the LLM acts as planner and orchestrator rather than the sole source of answers~\cite{yao2023react,schick2023toolformer,shinn2023reflexion,wu2023autogen}. This shift from a single model call to a tool-use loop lets LLMs engage with multi-step engineering workflows where text alone is not enough. Agentic systems can decompose a problem, orchestrate solvers and data sources, inspect results, and revise plans, but these skills have barely reached engineering workflows that demand verified, physically grounded computation.

Many power and energy systems tasks align closely with such an agentic workflow. Grid planning and operation, for example, combines language, data, physical models, optimization, and control in one procedure, where an operator may read a market manual, parse SCADA telemetry, run a power-flow solver, check a thermal limit, and develop a report. The data is heterogeneous, mixing textual artifacts such as grid codes, NERC reliability standards, outage reports, and vendor manuals with numerical streams from SCADA, weather forecasts, market signals, and asset records. The computational stack is also mixed, spanning forecasting models, power-flow and dynamic simulators, optimization solvers for economic dispatch (ED) and optimal power flow (OPF), and controllers for voltage and frequency regulation, each with its own interface and assumptions. The system must satisfy Kirchhoff's laws, thermal and voltage limits, ramp rates, and security margins, and a violation carries operational, economic, or safety consequences. Renewable forecasting, distributed energy resource (DER) coordination, EV charging scheduling, grid planning, contingency analysis, and demand response all sit in this kind of workflow, where 
LLM reasoning and agentic systems can act as the layer that ties these components together.

\looseness=-1 Power researchers have started to explore agentic AI's promises in power industry. Current exploration falls into three major realms. \textit{Capability studies} evaluate LLMs on various domain-specific tasks, e.g., power flow analysis, OPF, EV scheduling, forecasting, and retrieval-based question answering over technical knowledge bases~\cite{huang2024foundation,majumder2024electric}. \textit{Tool-augmented agents} place trusted solvers behind a language interface. GridMind couples LLMs with OPF and contingency solvers~\cite{jin2025gridmind}. Grid-Agent pairs a planning agent with a sandboxed validation agent to remediate operating-limit violations~\cite{zhang2025gridagent}. PowerGraph-LLM combines graph representations with in-context learning for OPF~\cite{bernier2025powergraphllm}. The PowerAgent roadmap integrates foundation models, the Model Context Protocol (MCP), and agentic workflows~\cite{zhang2025poweragent}. \textit{LLM-based forecasting} uses the LLM itself as the predictor, with promising results on wind power, load, and EV charging-station occupancy especially in data scarce settings~\cite{liu2025cpllmwpf,liao2026zerofewshot,lu2026evoccupancy}.

All three lines share common failure modes. LLMs hallucinate, generating numerically plausible answers with no basis in underlying systems~\cite{ji2023hallucination}, and have no built-in mechanisms to enforce physical laws such as power flow equations, line thermal limits, or generator ramp rates. A fluent dispatch report can therefore describe a schedule violating these constraints without flagging it~\cite{huang2024foundation,majumder2024electric}. LLM outputs are also hard to verify by inspection, since a correct numerical claim and an incorrect one are written with the same confidence and format, so an operator cannot judge credibility without re-running the trusted tool. This matters because grid outputs feed safety-critical control actions: a misreported voltage limit propagates into a dispatch decision, an under-delivered EV schedule becomes customer complaints and bill charges, and a missed contingency violation can damage equipment or cause blackouts. Power systems therefore need a deployment principle distinct from open-ended AI chatbots, one in which numerical claims are produced by trusted tools and gated by explicit verification before they reach the operator.

\begin{figure}[!ht]
    \centering
    \includegraphics[width=0.95\linewidth]{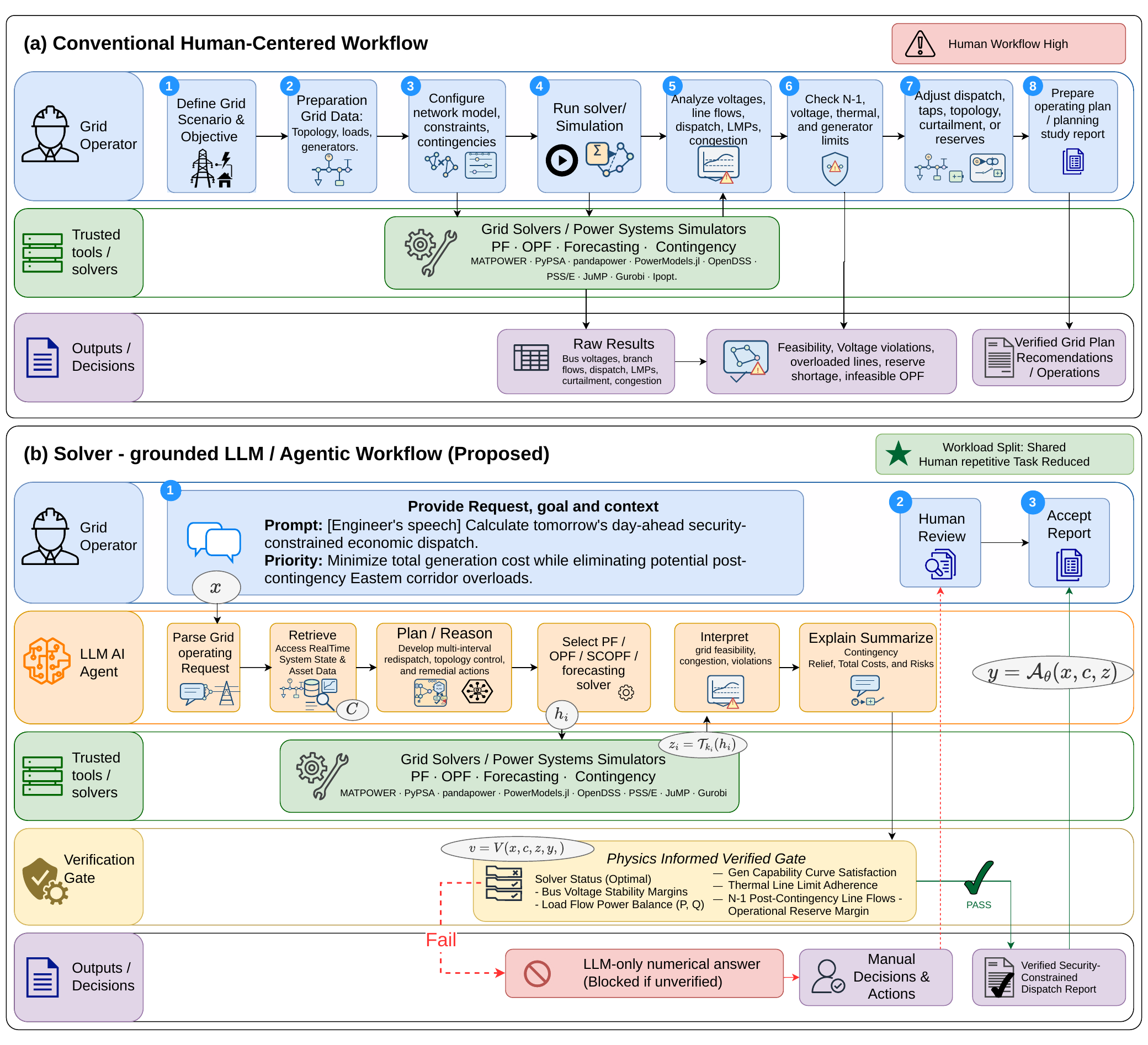}
    \caption{Solver-grounded LLM and agentic workflow for smart grids. (a) Conventional power system studies rely on trusted numerical tools, followed by explicit verification before results are reported. (b) In the proposed solver-grounded agentic workflow, the LLM acts as an interface and orchestration layer: it parses natural language requests, retrieves relevant context, plans tool calls, and summarizes outputs, but numerical quantities are reported only if they originate from a trusted solver or simulator and pass a verification gate. If verification fails, the solver-grounded workflow either repairs the setup through bounded retries, or returns a safe failure message rather than fabricating numerical results.}
    \label{fig:workflow_two_panel}
\end{figure}

This paper takes this principle as its central thesis: LLMs and agentic AI systems for smart grids should be designed as a solver-grounded orchestration layer. The proposed design separates three roles: The \textit{LLM} interprets the request, plans the task, calls tools, and generates the response. The \textit{trusted tool} performs the numerical computation. The \textit{verification gate} checks the result against physical and operational constraints before any value reaches the operator. Case studies in this paper span distinct smart grid task types, from forecasting and optimization to network analysis and diagnosis, and each is a setting where a plain LLM tends to produce plausible but infeasible numbers, so the solver-grounded design offers the clearest advantage. Fig.~\ref{fig:workflow_two_panel} illustrates this structure alongside the conventional power-system workflow, and we organize the rest of the paper around it. In summary, this paper makes three contributions:
\begin{enumerate}
\item \looseness=-1 We present a layered tutorial on LLM-based and agentic systems for smart grids, spanning prompting strategies (structured, few-shot, chain-of-thought, and RAG) and agentic architectures together with their engineering ingredients (typed tool interfaces, short- and long-term memory, and orchestration frameworks).
\item We formalize the solver-grounded design rule: a numerical result is reported only when it originates from a trusted tool and passes explicit verification. The rule excludes the LLM from computing physical quantities while keeping it in the closed loop. 
We instantiate it in four case studies (wind power forecasting, EV charging scheduling, power flow analysis, and contingency diagnosis). Each case study compares an LLM-only baseline against its solver-grounded counterpart on identical data and metrics, with reproducible results in Section~\ref{sec:casestudies}.
\item We propose a four-group evaluation framework covering task utility, solver-grounded correctness, faithfulness and safe failure, 
and cost and latency. All four groups are defined in Section~\ref{sec:solver-grounded}. The framework supports head-to-head comparison across architectures and surfaces failure modes that task utility alone would hide.
\end{enumerate}

For power and energy researchers, grid operators and engineers, graduate students, and AI practitioners, this work guides both what LLMs can achieve for smart grids and how to build such systems responsibly and evaluate them under explicit protocols. We treat the paper as a tutorial rather than a pure literature review, as it lays out concepts, design rules, architectural patterns, and worked examples, so the reader leaves with the vocabulary and understanding to design these systems and the building blocks to develop them.

The rest of the paper is organized as follows. Section~\ref{sec:related} reviews related work. Section~\ref{sec:workflow} covers LLM preliminaries and formalizes the solver-grounded design rule. Section~\ref{sec:prompting} covers prompting systems (structured, few-shot, and chain-of-thought prompting, and RAG), and Section~\ref{sec:agents} covers agentic architectures (ReAct, plan-and-act, hierarchical, self-reflection, and multi-agent) with memory and engineering frameworks. Section~\ref{sec:evaluation} details the evaluation protocol, Section~\ref{sec:casestudies} presents the four case studies that instantiate it, and Section~\ref{sec:conclusion} concludes with open problems: verification design, repair policies, agentic benchmarks, and safer deployment.

\subsection{Related works}
\label{sec:related}

\looseness=-1 Literature on LLMs and agentic AI for smart grids have grown since 2023, and now spans four directions: (i) domain surveys, (ii) LLM-based forecasting and time-series modeling that use LLMs as prediction models, (iii) tool-augmented agents that wrap trusted solvers behind a language interface, and (iv) LLM-driven planning, optimization, and operations agents that embed LLMs in closed-loop workflows for grid operation and DERs. We review representative works in each and discuss how they motivate this tutorial.

\begin{table}[!ht]
\centering
\footnotesize
\caption{Summary of reviewed literature on LLMs and agentic AI systems for smart grids.}
\label{tab:related_work_summary}
\begin{tabular}{p{0.1\textwidth} p{0.35\textwidth} p{0.19\textwidth} p{0.25\textwidth}}
\hline
\textbf{Literature stream} & \textbf{Main techniques} & \textbf{Typical grounding \newline mechanism} & \textbf{Remaining gap} \\
\hline
Domain \newline surveys & Map LLM capabilities, risks, applications, and deployment scenarios across power-system tasks~\cite{huang2024foundation,majumder2024electric,amjad2025review,sarwar2025survey,madani2025smartgrids} & Literature-level analysis and task taxonomies & Limited guidance on how to enforce numerical correctness and physical feasibility in deployed systems \\
\hline
LLM-based forecasting and time series modeling & Adapt LLMs to load, renewable-generation, household-energy, and EV-occupancy prediction tasks~\cite{tang2025timeseries,wu2024timeseries_llm,takrouri2025distillation,liao2026zerofewshot,liu2025cpllmwpf,lu2026evoccupancy} & Prompting, fine-tuning, distillation, or multimodal temporal representations & Forecasts may improve generalization or usability, but they are not solver-grounded and require task-specific validation \\
\hline
Tool-augmented agents & Couple LLMs with simulators, optimizers, validation agents, and domain tools for OPF, contingency analysis, violation detection, remediation, and grid-model generation~\cite{jin2025gridmind,zhang2025gridagent,wen2025xgridagent,jia2025feedback,zhang2025poweragent,deng2025gridmodel,liu2025repower,saha2025dragent} & External solvers, simulators, sandboxed validation, and structured tool calls & Existing systems are task-specific and lack a shared evaluation protocol or general correctness rule \\
\hline
Planning, optimization, and operations agents & Embed LLMs in dispatch, unit commitment, building energy management, demand response, EV scheduling, reliability monitoring, and infrastructure planning workflows~\cite{hu2025solverready_uc,cheng2025gaia,mohammadi2025ed,liu2025dispatch_redesign,jena2025incontext_mpc,makroum2025hems,ji2025lemad,cheng2026rag} & Solver-ready formulations, optimization tools, retrieval, knowledge graphs, or multi-agent coordination & The boundary between LLM reasoning, trusted numerical computation, and verification is often left implicit \\
\hline
\end{tabular}
\end{table}

\textbf{Domain surveys.}
The first stream assesses the capabilities and limits of current LLMs across power-system applications, from OPF, EV scheduling, and power-flow analysis to retrieval-augmented question answering, consistently identifying a gap between fluent text and physically grounded computation~\cite{huang2024foundation,majumder2024electric,hamann2024grid_fms,ibrahim2025cyber_survey}. A second group builds task-oriented taxonomies spanning fault diagnosis, forecasting, control, grid operations, markets, and security, and reviews agentic AI architectures with their tool-injection and governance risks~\cite{amjad2025review,sarwar2025survey,madani2025smartgrids,agentic_power_2025,huang2024llm_optimization_survey,xie2026foundation}. These studies establish what current LLMs can and cannot do, but most do not specify how to build systems that close the capability-correctness gap.

\textbf{LLM-based forecasting and time-series modeling.}
A second stream applies LLMs directly as numerical predictors in energy systems, exploiting large-scale pretraining and prompting or fine-tuning to transfer across sites and data-scarce settings~\cite{tang2025timeseries,wu2024timeseries_llm}. Representative work spans knowledge distillation and zero-shot load forecasting, multi-scale prompting for wind power, and explainable frameworks for EV charging-station occupancy~\cite{takrouri2025distillation,liao2026zerofewshot,liu2025cpllmwpf,lu2026evoccupancy}. Our wind-forecasting case study (Section~\ref{sec:casestudies}-A) complements these works with a head-to-head comparison of prompting strategies against a GRU baseline.

\textbf{Tool-augmented agents.}
A third stream wraps trusted solvers behind a language interface, delegating computation to simulators, optimizers, and validation routines rather than trusting the LLM's own numbers. Systems couple LLMs with simulators such as pandapower through multi-agent, validation-based, or hierarchical architectures for AC OPF, N-1 contingency analysis, and autonomous remediation of operating-limit violations, routing simulator error signals back to planning agents or organizing the design space around foundation models and MCP~\cite{jin2025gridmind,zhang2025gridagent,wen2025xgridagent,jia2025feedback,zhang2025poweragent,bernier2025powergraphllm,deng2025gridmodel,liu2025repower,saha2025dragent}. They match or approach solver performance on individual tasks but share no common evaluation protocol or correctness principle, the gap we address in Sections~\ref{sec:solver-grounded} and~\ref{sec:casestudies}.

\textbf{LLM-driven planning, optimization, and operations agents.}
A fourth stream embeds LLMs in closed-loop planning, optimization, and control workflows for grid operation and DERs. For dispatch and optimization, prior work translates natural-language descriptions into solver-ready unit-commitment models, adapts LLMs for power dispatch, and combines them with knowledge graphs and agent layers~\cite{hu2025solverready_uc,cheng2025gaia,mohammadi2025ed,liu2025dispatch_redesign}. Other work embeds LLMs as topology-adaptive MPC surrogates, coordinates multi-appliance scheduling against MILP baselines, supports reliability monitoring through multi-agent and retrieval-augmented systems, and addresses EV behavior modeling, demand response, and charging-infrastructure planning~\cite{jena2025incontext_mpc,makroum2025hems,ji2025lemad,cheng2026rag,xiong2026evbehavior,zhang2025drev,zheng2025evplanning}. Our EV-scheduling case study (Section~\ref{sec:casestudies}-B) contributes an agentic system that calls a CVXPY solver, compared against a standalone LLM baseline on the Caltech ACN-Data benchmark~\cite{lee2019acndata}.

\section{Preliminaries and the Solver-Grounded Design Rule}
\label{sec:workflow}
This section provides background the rest of the paper builds on. We review LLMs as probabilistic sequence models (Section~\ref{sec:llm-basics}), summarize frontier LLMs and their training procedures (Section~\ref{sec:frontier-llms}), formalize the \emph{solver-grounded design rule}, the paper's central concept (Section~\ref{sec:solver-grounded}),
and identify workflow bottlenecks where LLMs and agents add the most value under it (Section~\ref{sec:where-llms-help}).

\subsection{Large Language Models and Next-Token Prediction}
\label{sec:llm-basics}

Large language models are probabilistic autoregressive sequence models, trained to predict the next token given previous tokens~\cite{brown2020language}. Formally, for a token sequence $w_{1:T}$, the model defines token probability as
\begin{equation}
p(w_{1:T}) = \prod_{t=1}^{T} p(w_t \mid w_{<t}),
\label{eq:lm_factorization}
\end{equation}
where $T$ is the sequence length and $w_{<t} = (w_1, \dots, w_{t-1})$ denotes the tokens preceding position $t$. During pretraining, model parameters $\theta$ are optimized on a large text corpus $\mathcal{D}$ by minimizing the negative log-likelihood
\begin{equation}
\label{loss_predict}
\mathcal{L}_{\text{pre}}(\theta)
= - \sum_{w \in \mathcal{D}} \sum_{t=1}^{T}
\log p_{\theta}(w_t \mid w_{<t}),
\end{equation}
which is equivalent to maximizing the likelihood of the training sequences. 

The next-token-prediction objective in~(\ref{eq:lm_factorization})--(\ref{loss_predict}) has two implications for LLM-based smart grid systems. First, since generation is conditioned on the available context $w_{<t}$, output quality depends strongly on the information provided at inference time. This motivates the prompting and context-construction methods discussed later, including structured, few-shot, and chain-of-thought prompting and RAG, which improve generation by supplying task descriptions, examples, reasoning structure, or external domain knowledge. Second, next-token prediction does not by itself enforce physical, operational, or numerical correctness. LLMs are trained to model patterns in large text corpora, not to guarantee physical laws such as Kirchhoff's laws or optimization feasibility in power grids. The model also has no intrinsic way to distinguish quantities computed by a trusted solver from quantities inferred through language generation, and both appear equally fluent in the output. This explains why an LLM can produce a plausible but infeasible dispatch or an incorrect power flow result, and motivates the solver-grounded design rule of Section~\ref{sec:solver-grounded}, where correctness is enforced with trusted tools and verification.

\subsection{Frontier LLM Families and Training Procedures}
\label{sec:frontier-llms}

Frontier LLMs split into \emph{proprietary} families accessed through managed APIs and \emph{open-weight} families deployable locally, the latter suitable when operational data is not allowed to be accessed out of the utility's network~\cite{openai_gpt55_2026,anthropic_opus48,google_gemini31pro_2026,meta_llama4scout_2026,deepseek_r1_2025,qwen_qwen3instruct2507_2025,mistral_large3_2025}. Across both, context length, multimodal-input support, and structured-output reliability vary by family and model size, and the model landscape evolves on a fast timescale of months.

The frontier models share a common three-stage training pipeline: self-supervised pretraining on raw text, supervised fine-tuning on instruction-response data, and preference-based alignment.

In the \textit{pretraining stage}, the model learns broad statistical patterns of language by minimizing the next-token loss in~(\ref{loss_predict}) over a large unlabeled corpus. This gives general linguistic competence and world knowledge but targets no specific task or instruction format, so a pretrained-only model continues text fluently yet may not reliably follow instructions, answer in the expected format, or refuse unsafe requests.

\textit{Supervised fine-tuning (SFT)} then adapts the pretrained model to instruction following using curated prompt-response demonstrations~\cite{wei2022finetuned}, updating the parameters by minimizing the token-level cross-entropy between predicted and target responses. SFT lets LLMs give structured, formatted answers and is a prerequisite for the tool calling, structured outputs, and conversational behavior agentic systems rely on.

A final alignment stage optimizes the SFT model with human preference feedback so responses better match expectations for helpfulness, safety, and reliability~\cite{ouyang2022training}. The most common implementation is \textit{reinforcement learning from human feedback (RLHF)}, where a reward model trained on human preference comparisons is used to optimize the LLM through reinforcement learning. A related approach, \textit{reinforcement learning from verifiable rewards (RLVR)}, replaces the learned reward model with an automatic verifier that scores each response against a ground-truth criterion
~\cite{lambert2024tulu3}. Note that alignment improves how the model behaves, but does not by itself ensure numerical correctness or constraint satisfaction, so in smart grid workflows it should be treated as a behavior layer on top of trusted solver and verification components.

\subsection{The Solver-grounded Design Rule}
\label{sec:solver-grounded}

The central design rule of this paper is that LLMs and agents should not be treated as sources of numerical truth in smart grid applications. Instead they act as interfaces, task planners, and orchestrators around trusted computational tools. These tools are central to safety-critical power systems practice, such as operating and planning studies already rely on power-flow solvers, OPF solvers, economic-dispatch optimizers, forecasting models, and simulators to enforce physical laws, operational limits, and feasibility. In current workflows, engineers configure cases, run these tools, inspect convergence and constraint violations, iterate across scenarios, and translate the outputs into operational reports. 

LLMs fit tool calling naturally because they can map natural language requests into structured tool inputs, select and sequence the right solver or data API, interpret logs and outputs, and explain results while leaving numerical computation to verified softwares. A numerical result should therefore be reported only if a trusted solver or simulator produces it and it passes explicit verification. This separates three roles. The LLM interprets the request, decomposes the task, selects tools, and explains results. The trusted tool performs the numerical computation. The verification gate checks whether the output is physically and operationally valid. Formally, let $\{\mathcal{T}_k\}_{k=1}^{K}$ denote a set of trusted tools, such as power flow (PF) solvers, OPF solvers, ED optimizers, forecasting models, or domain-specific simulators. During an agentic workflow, the system may call one or more tools on intermediate task states $h_i$, producing a collection of tool outputs 
\begin{equation}
z_i = \mathcal{T}_{k_i}(h_i),
\end{equation}
where $h_i$ is the intermediate task state at step $i$ and $k_i \in \{1,\ldots,K\}$ indexes the tool the agent selects. The state $h_i$ is a workflow-level object. It holds the current task context, such as the parsed request, the network state, and prior observations. It is distinct from the token-level sequence $w_t$ of Section~\ref{sec:llm-basics}. The LLM or agent $\mathcal{A}_{\theta}$ then generates a candidate response
\begin{equation}
y = \mathcal{A}_{\theta}(x,c,z),
\end{equation}
where $x$ is the user request or original task input, $c$ denotes additional context such as retrieved documents, prompt instructions, memory, or intermediate plans, and $z = \{z_i\}$ is the collection of tool outputs generated during the workflow, and $y$ is the candidate response produced by the LLM or agent.

{The final response is evaluated by a verification function that operates at the task-output level:
\begin{equation}
v = \mathcal{V}(x,c,z,y) \in \{\text{pass}, \text{fail}\} \times \mathbb{R}^m,
\label{eq:verification}
\end{equation}
Here $m$ is the number of task-specific verification conditions. The first component of $v$ is a binary status flag. The second is a vector of the $m$ residuals, one per condition. These conditions are of two kinds. \emph{Constraint-satisfaction} checks (e.g., maximum voltage violation, total energy shortfall, or active-power balance error) test physical and operational feasibility. \emph{Faithfulness} checks test whether each numerical value reported in $y$ matches its trusted-tool value in $z$. Each condition passes when its residual lies within a task-specific tolerance, and the status flag is $\text{pass}$ only when all conditions pass. Section~\ref{sec:evaluation} specifies these checks and their tolerances.

{We call an LLM or agentic system \textbf{solver-grounded} if its final response $y$ passes the task-level verification in~(\ref{eq:verification}). Every reported numerical result must be traceable to trusted tool evidence in $z$, and the final response must satisfy all the verification conditions. When the task-level verification condition is not met, the system should explicitly report the failure mode rather than fabricate, extrapolate, or silently repair the answer. We refer to this behavior as \textbf{safe failure}. It adapts abstention from selective prediction~\cite{geifman2017selective} to the tool-use setting.  
}

{This rule turns correctness from a property of the language model alone into a property of the overall LLM tool verification agentic system. 
A solver-grounded agentic system reports only numerical claims that are tool-derived, verifiable, and faithfully represented in the final response. 
} Fig.~\ref{fig:workflow_two_panel} contrasts the power system workflow with its agent-assisted counterpart. The verification gate is identical in both panels. What changes on the right is who calls the solver, reads violation messages, and decides alternatives to retry. The rule is conservative by construction: it leaves room for LLMs in any non-numerical step, such as parsing requests, retrieving context, drafting summaries, and explaining solver logs, but excludes them from computing physical quantities. Section~\ref{sec:evaluation} presents the evaluation protocol and verification process.

One boundary deserves explicit statement. A \emph{trusted tool} here does not mean an infallible source of truth. Instead, such a tool refers to a model, solver, or simulator whose assumptions are explicit and whose outputs can be checked for convergence, feasibility, and consistency with power-system constraints. This matches current practice: engineers rely on power-flow solvers, OPF solvers, scheduling optimizers, and contingency simulators, but still verify solver status, constraint residuals, voltage and line-flow limits, and input assumptions before using the results operationally.

Such a principle applies most directly when such a tool exists for the reported quantity, as in the EV scheduling, power flow, and contingency case studies, where a solver computes the result and the verification gate checks its validity for the specified model and data. When no trusted solver exists, as in LLM-based wind power forecasting, the output cannot be solver-grounded in the same sense, and verification is limited to model- and output-level checks such as schema validity, range feasibility, consistency with historical data, and forecast-error evaluation against observations. This is the boundary of the solver-grounded regime rather than an exception: absent a trusted tool, the system should report predictions with their validation evidence and uncertainty, not as solver-verified quantities.

\subsection{Where LLMs and Agents Add the Most Value}
\label{sec:where-llms-help}

The solver-grounded rule of Section~\ref{sec:solver-grounded} guardrails what LLMs should not do in smart grid tasks. 
In practice, there are several recurring bottlenecks in solver-grounded smart grid workflows: data preparation, model specification, tool-run configuration, failure detection, scenario iteration, and technical reporting and interpretation. We do not propose these as a formal taxonomy but as a high-level synthesis of workflow patterns that motivate the observations and case studies in Section~\ref{sec:casestudies}.
 
\emph{First, bottlenecks live around rather than inside the solver.} Data preparation, model setup, iteration, and reporting consume most expert time per scenario, while the solver call is fast. This supports positioning the LLM as an interface and orchestration layer rather than a numerical engine. The PFAgent case study (Section~\ref{sec:pfagent}) shows the split: PandaPower's Newton-Raphson solver computes the operating point, while the LLM parses multi-step queries, selects and sequences tools, tracks network state across turns, and synthesizes outputs into operator-readable responses.

\emph{Second, iteration cost compounds with case complexity.} As grid studies grow from one case to many scenarios, contingencies, assets, and operating conditions, the burden shifts from solving one well-posed problem to repeatedly specifying tasks, selecting tools, checking inputs, interpreting outputs, and revising the workflow. Here LLMs and agents add value as planners and orchestrators, since they decompose a request into subtasks, identify the required data and tools, sequence tool calls, and adapt to intermediate results, leaving numerical computation to trusted solvers. The GridDebugAgent case study (Section~\ref{sec:griddebug}) shows this by automating the diagnose-act-verify loop while preserving solver-grounded correctness.
 
\emph{Third, verification itself is a bottleneck.} The final output often combines numerical results, explanations, diagnostics, and recommendations. Even when the solver call is correct, the agent can misreport a value, omit a violation, or overstate a result's validity. A solver-grounded design therefore separates three functions: i). the LLM interprets, plans, orchestrates, and explains, ii). trusted tools produce the numerical evidence, and iii). a verification gate decides whether the final answer is safe to communicate. Case studies quantify these roles, where EVAgent produces zero hard physical constraint violation across 20 benchmark days, whereas a standalone LLM baseline violates the 50\,kW site cap.

\section{Prompting Strategies for Smart Grids}
\label{sec:prompting}

We start with prompting-based systems, where the LLM produces the output directly from its input context, without external tool calls. We cover four strategies: structured, few-shot, and chain-of-thought prompting, and RAG. Fig.~\ref{fig:prompting_methods} compares them. Prompting is the simplest design layer and the baseline against which we compare the agentic systems of Section~\ref{sec:agents}.

\subsection{Structured Prompting}
\label{sec:structured-prompting}
Structured prompting normally decomposes the prompt into four parts~\cite{liu2023pretrain,schulhoff2024prompt}: the \emph{role} defines the domain persona, the \emph{context} gives the problem setup, assumptions, and constraints, the \emph{task} states the instruction, and the \emph{output format} specifies the required fields and units. Fig.~\ref{fig:structured_prompt_example} illustrates such a prompt on a textbook power-factor problem. The model reports active power, apparent power, and power factor correctly but hallucinates the total reactive power, \footnote{The prompting examples in Figs.~\ref{fig:structured_prompt_example}--\ref{fig:rag_prompt_example} were generated with GPT-4o via the OpenAI API (April 2026).} 
returning 100 var where its own active and apparent power imply $Q=\sqrt{S^2-P^2}\approx 154.7$ var, an error a reader cannot catch without recomputing. Structured prompting thus guides the model toward a well-organized response, while it could not self-guarantee numerical correctness.

In smart grid applications, structured prompting helps replace an open-ended conversation with a controlled interaction.
For an LLM summarizing an OPF or ED study, the role sets an operator persona, the context supplies the setup and physical constraints, the task requests the summary and any limit violations, and the output section fixes a table or JSON schema with prescribed fields and units. The result is parseable by downstream tools or a verification gate, though still not guaranteed numerically correct. Such prompts have been used to specify economic-dispatch and unit-commitment problems for LLM-based solving~\cite{mohammadi2025ed,hu2025solverready_uc}.

\begin{figure}[!ht]
\centering
\begin{minipage}{0.94\linewidth}
\begin{promptfigure}{Example of Structured Prompting}
\scriptsize
\noindent\textbf{Prompt input}\hspace{1.2mm}%
\legendchip[colback=rolebg]{Role}\hspace{0.7mm}%
\legendchip[colback=contextbg]{Context}\hspace{0.7mm}%
\legendchip[colback=taskbg]{Task}\hspace{0.7mm}%
\legendchip[colback=outputbg]{Output format}\par\vspace{0.8mm}

{\ttfamily
\rolehl{You are a power systems assistant. Use only the given data and standard power system definitions.}
\contexthl{Three loads are connected in parallel: L1 = 200 W at unity power factor, L2 = 400 W at 0.8 lagging, and L3 = 300 W at 0.9 leading.}
\taskhl{Compute the total active power, reactive power, apparent power, and overall power factor of the aggregated load.}
\outputhl{Return the answer in four fields: total active power, total reactive power, total apparent power, and overall power factor.}
}

\tcblower

\scriptsize
\textbf{Representative model output}\par\vspace{0.6mm}
\outputline{\goodmark\ Total active power: 900 W}
\outputline{\badmark\ Total reactive power: 100 var (lagging)}
\outputline{\goodmark\ Total apparent power: 913.2 VA}
\outputline{\goodmark\ Overall power factor: 0.9855 lagging}
\end{promptfigure}
\end{minipage}
\caption{Structured prompting example (role, context, task, output format) based on Example 2.5 (\textit{Parallel Loads}) from \cite{kirschen2024power}. The model predicts total active power, apparent power, and power factor correctly but hallucinates the total reactive power, reporting 100 var against the 154.7 var implied by its own apparent power and power factor.} 
\label{fig:structured_prompt_example}
\end{figure}

\subsection{Few-Shot Prompting}
Few-shot prompting conditions the output on a few input-output demonstrations~\cite{brown2020language}. Each demonstration is a solved instance pairing an example input with its correct output, placed in the prompt just before the target query. From these examples the model infers the intended input-output mapping and answer format and responds to the new query by analogy, with no parameter update, a behavior known as in-context learning. Even very few demonstrations help improve formatting consistency and task accuracy over a plain instruction. Fig.~\ref{fig:fewshot_prompt_example} shows few-shot prompting on the same power factor problem with two demonstrations.

\begin{figure}[ht]
\centering
\begin{minipage}{0.94\linewidth}
\begin{promptfigure}{Example of Few-Shot Prompting}
\scriptsize
\noindent\textbf{Prompt input}\hspace{1.2mm}%
\legendchip[colback=exonebg]{Example 1}\hspace{0.7mm}%
\legendchip[colback=extwobg]{Example 2}\hspace{0.7mm}%
\legendchip[colback=targetbg]{Target task}\hspace{0.7mm}%
\legendchip[colback=outputbg]{Output format}\par\vspace{0.8mm}

{\ttfamily
You are a power systems assistant.
\exonehl{Load A has active power 100 W at unity power factor. Return total active power, reactive power, apparent power, and power factor. Answer: 100 W, 0 var, 100 VA, unity power factor. }
\extwohl{Load B has active power 200 W at 0.8 lagging. Return total active power, reactive power, apparent power, and power factor. Answer: 200 W, 150 var, 250 VA, 0.8 lagging. }
\targethl{Now solve this problem: three loads are connected in parallel: L1 = 200 W at unity power factor, L2 = 400 W at 0.8 lagging, and L3 = 300 W at 0.9 leading. Compute the total active power, reactive power, apparent power, and overall power factor. }
\outputhl{Return the answer in four fields: total active power, total reactive power, total apparent power, and overall power factor.}
}

\tcblower

\scriptsize
\textbf{Representative model output}\par\vspace{0.6mm}
\outputline{\goodmark\ Total active power: 900 W}
\outputline{\goodmark\ Total reactive power: 154.7 var (lagging)}
\outputline{\goodmark\ Total apparent power: 913.2 VA}
\outputline{\goodmark\ Overall power factor: 0.9855 lagging}
\end{promptfigure}
\end{minipage}
\caption{Few-shot prompting example for the same textbook-style power systems problem. Colors indicate example demonstrations, the target task, and the requested output format.}
\label{fig:fewshot_prompt_example}
\end{figure}

In smart grid applications, few-shot prompting teaches domain conventions and the expected input-output relationship. In day-ahead wind forecasting, for instance, demonstrations can supply historical wind power data and show how previous days shape the next day's output. Two limitations matter: demonstrations consume more prompt budget than a plain instruction, and the method is sensitive to example choice, biasing the model toward the examples' assumptions and format even when the target case differs. Examples should therefore cover representative task variations, and outputs should still be verified by downstream checks. In power systems, few-shot and in-context prompting have been applied to load forecasting and OPF prediction~\cite{liao2026zerofewshot,bernier2025powergraphllm}.

\subsection{Chain-of-thought Prompting}
Chain-of-thought (CoT) prompting asks the model to produce intermediate reasoning steps before the final answer~\cite{wei2022chain}. Because generation is sequential, the steps it writes become context for the final answer, so it decomposes a multi-step problem into sub-steps and builds the answer on its own reasoning. Making the steps explicit lets a reader follow how the answer was reached and improves quality on multi-step tasks. Fig.~\ref{fig:cot_prompt_example} illustrates CoT prompting on a power flow over-voltage diagnosis.

In smart grid applications, CoT helps when the LLM explains solver outputs or supports human review. For a bus over-voltage, a CoT response can identify the violated limit, locate the affected bus, connect it to causes such as excessive local generation, 
and suggest actions such as adjusting inverter setpoints or transformer taps. 
Like the other techniques, CoT does not guarantee correct reasoning. It can produce plausible reasoning while making an arithmetic error or recommending an infeasible action, so the final response must still be checked against trusted computations and constraints. In power systems, CoT has been used within LLM assistants for dispatch operations~\cite{cheng2025gaia}.

\begin{figure}[!ht]
\centering
\begin{minipage}{0.94\linewidth}
\begin{promptfigure}{Example of Chain-of-Thought Prompting}
\scriptsize
\raggedright
\noindent\textbf{Prompt input}\hspace{1.2mm}%
\legendchip[colback=contextbg]{Problem}\hspace{0.7mm}%
\legendchip[colback=reasonbg]{Reasoning cue}\hspace{0.7mm}%
\legendchip[colback=outputbg]{Output format}\par\vspace{0.8mm}

{\ttfamily\raggedright
You are a power systems assistant.\par
\contexthl{A power flow solution on a distribution feeder reports the voltage at bus 7 as 1.063 p.u. The allowed range is 0.95--1.05 p.u. The feeder is lightly loaded at midday with high rooftop PV generation near bus 7.}\par
\reasonhl{Diagnose the result step by step. First check the reported voltage against its limit. Then identify the most likely physical cause. Finally propose a corrective action.}\par
\outputhl{Return both the intermediate reasoning steps and the final diagnosis with a recommended action.}
}

\tcblower

\scriptsize
\textbf{Representative model output}\par\vspace{0.6mm}
\outputline{\goodmark\ Step 1: 1.063 p.u. exceeds the 1.05 p.u. upper limit, so bus 7 is in over-voltage.}
\outputline{\goodmark\ Step 2: Light load with high local PV pushes real power back toward the substation, raising the local voltage.}
\outputline{\goodmark\ Step 3: Reduce the net injection or absorb reactive power near bus 7.}
\outputline{\goodmark\ Diagnosis: over-voltage at bus 7 (1.063 vs.\ 1.05 p.u.) from reverse power flow under light load; action: command the PV inverters near bus 7 to absorb reactive power, or lower the transformer tap.}
\end{promptfigure}
\end{minipage}
\caption{Chain-of-thought prompting example on a power flow over-voltage diagnosis. The reasoning cue makes the model lay out the limit check, the likely physical cause, and a corrective action step by step, rather than returning a one-line verdict.}
\label{fig:cot_prompt_example}
\end{figure}

\subsection{Retrieval-augmented Generation (RAG)}

RAG conditions the output on external documents retrieved at inference time~\cite{lewis2020rag}. An LLM answering from its internalized memory alone can produce plausible but outdated or unsupported statements. RAG addresses this in two stages. A retriever first searches an external knowledge source for the passages most relevant to the query. Those passages are inserted into the prompt as context, so the model answers from the retrieved evidence rather than its parameters alone, often citing the passages used. In practice, source documents are split into self-contained chunks, retrieved with metadata filters such as jurisdiction, voltage level, or effective date, and reranked for relevance. The prompt must return an explicit ``insufficient evidence'' response when the passages do not support an answer. Fig.~\ref{fig:rag_prompt_example} illustrates the pipeline on a grid-code compliance question.

In smart grid applications, RAG helps because many task-relevant facts are not reliably stored in model parameters, including operating procedures, grid codes, market rules, reliability standards, tariff documents, and equipment manuals.
Consider a question about whether a proposed operating action is allowed under a specific interconnection rule, market manual, or utility procedure. A RAG system first retrieves the relevant document sections, such as the applicable voltage range, effective date, or operating condition. The prompt then asks the model to summarize the rule, decide whether the action satisfies it, and cite the supporting passage. This matters most when the answer depends on local rules or current documentation rather than general power-system knowledge. In power systems, RAG has grounded operational-reliability assessment in reliability documentation and operating procedures~\cite{cheng2026rag}.

\begin{figure}[ht]
\centering
\begin{minipage}{0.94\linewidth}
\begin{promptfigure}{Example of Retrieval-Augmented Generation}
\scriptsize

\textbf{Step 1: Retrieval}\par\vspace{0.6mm}
\legendchip[colback=querybg]{Retriever query}\hspace{0.7mm}%
\legendchip[colback=retrievedbg]{Top retrieved chunk}\par\vspace{0.8mm}
{\ttfamily
\queryhl{steady-state voltage limit point of common coupling sustained over-voltage curtailment}
\retrievedhl{Top retrieved chunk: Grid Code Sec.~7.3.1. At the point of common coupling, steady-state voltage shall be kept within 0.95--1.05 p.u. A sustained excursion above 1.05 p.u. for more than 60 s requires the operator to curtail local generation.}
} {\scriptsize\itshape Source: Distribution Grid Code, Sec.~7.3.1, chunk \#12}\\
\noindent\textbf{Step 2: Augmented prompt passed to the LLM}\par
\hspace{1.2mm}%
\legendchip[colback=retrievedbg]{Retrieved context}\hspace{0.7mm}%
\legendchip[colback=querybg]{Query}\hspace{0.7mm}%
\legendchip[colback=taskbg]{Task}\hspace{0.7mm}%
\legendchip[colback=outputbg]{Output format}\par
{\ttfamily
You are a power systems assistant.
\retrievedhl{Retrieved context: Grid Code Sec.~7.3.1. At the point of common coupling, steady-state voltage shall be kept within 0.95--1.05 p.u. A sustained excursion above 1.05 p.u. for more than 60 s requires the operator to curtail local generation. }
\queryhl{A feeder bus at the point of common coupling has been at 1.062 p.u. for 90 s due to midday PV. What does the grid code require? }
\taskhl{Using only the retrieved context, state whether action is required and which action, and cite the supporting clause. }
\outputhl{Return the required action and the cited clause.}
}
\tcblower
\scriptsize
\textbf{Representative model output}\par\vspace{0.6mm}
\outputline{\goodmark\ 1.062 p.u. exceeds the 1.05 p.u. steady-state limit and has persisted for 90 s ($>$60 s).}
\outputline{\goodmark\ Required action: the operator must curtail local generation.\goodmark\ Source: Grid Code Sec.~7.3.1.}
\end{promptfigure}
\end{minipage}
\caption{Retrieval-augmented generation example on a grid-code compliance question. The retriever returns the relevant clause with its source, and the model answers from the retrieved text and cites it, rather than relying on parametric knowledge of the code.}
\label{fig:rag_prompt_example}
\end{figure}

We treat RAG as a prompting method because it changes only the context provided to the model, not its parameters or its access to external computation. When retrieval is instead combined with tool execution, iterative planning, and verification, the resulting system becomes agentic, the subject of Section~\ref{sec:agents}.

\begin{figure}[tb]
    \centering
    \includegraphics[width=0.95\linewidth]{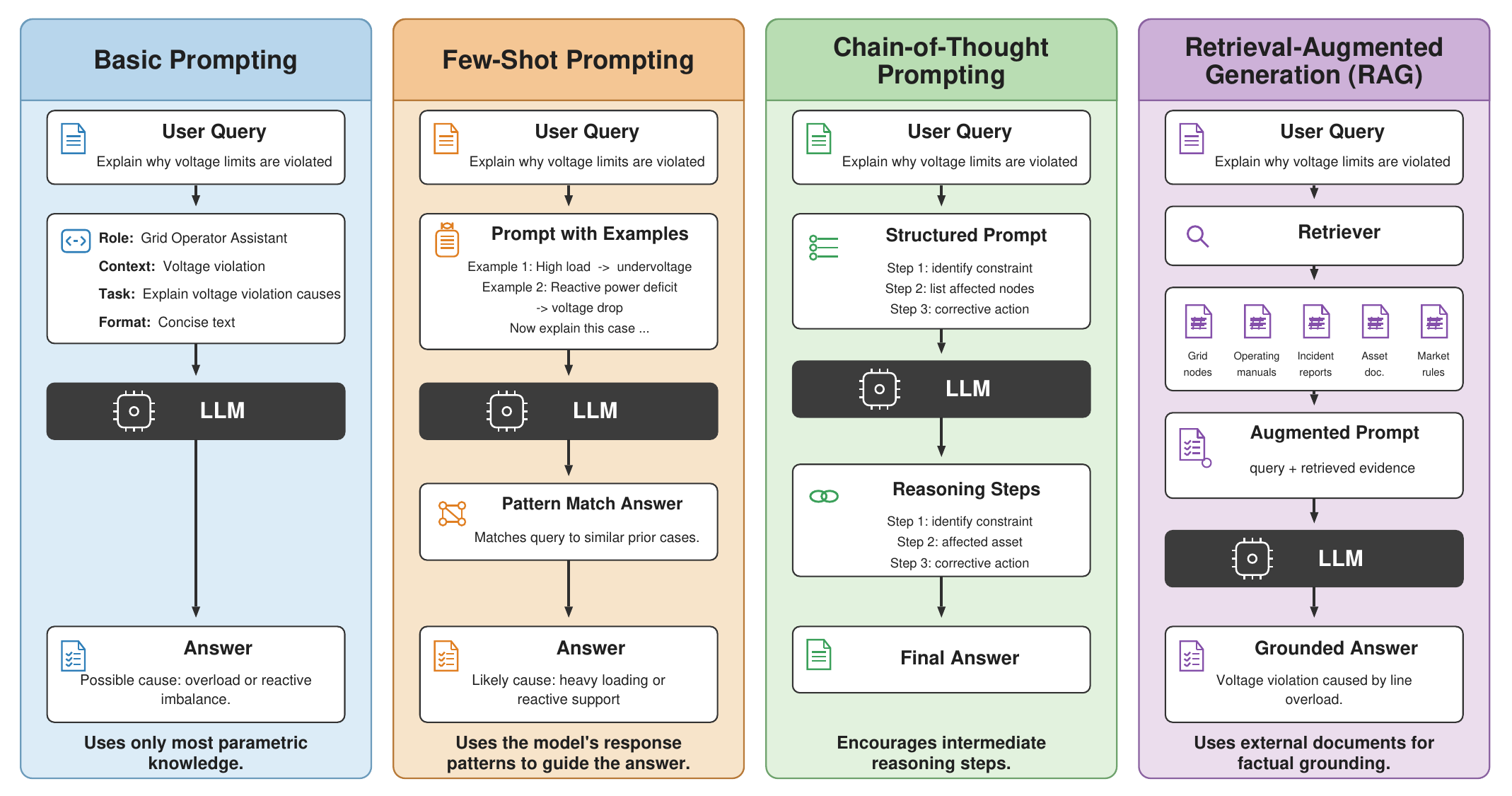}
    \caption{Comparison of representative prompting methods for smart grid tasks. Structured prompting relies on prompt design alone, few-shot adds example demonstrations, chain-of-thought requests intermediate reasoning steps, and RAG grounds the response in retrieved documents. }
    \label{fig:prompting_methods}
\end{figure}

\section{Agentic Systems for Smart Grids}
\label{sec:agents}

Moving from prompting to agentic systems, the LLM is no longer restricted to its input context but orchestrates external tools, inspects intermediate results, and decides what to do next. We first define an agentic system and describe how it acts on the environment through API calls and tool use. We then cover five representative architectures: Reasoning and Acting (ReAct), plan-and-act, hierarchical, self-reflection, and multi-agent systems. We close with two cross-cutting components every architecture relies on: memory, which preserves state across steps, and the frameworks used to build agents in practice.

\subsection{Agentic System Definition}
\label{sec:agent-def}
An agentic system extends the LLM beyond a single forward pass: instead of producing the answer directly from its input context, it solves a task through a sequence of reasoning steps and actions, where each action can query an external tool and return a new observation~\cite{yao2023react}. A common representation is the ReAct trajectory
\begin{equation}
\tau_t = (x, \text{Thought}_1, \text{Action}_1, 
\text{Observation}_1, \dots, \text{Thought}_t, 
\text{Action}_t, \text{Observation}_t),
\end{equation}
where $x$ is the initial user input or task description and $\tau_t$ is the partial trajectory up to step $t$. At each step, $\text{Action}_t \in \mathcal{A}$ is drawn from an action space $\mathcal{A}$ that includes LLM reasoning steps, such as decomposing the task or drafting a candidate answer, and external tool invocations, such as calling a power flow solver, querying a database, or executing a script. The resulting $\text{Observation}_t \in \mathcal{O}$ may come from an external tool, the operating environment, or the LLM's own reasoning traces when intermediate thoughts and self-evaluations are fed back into later steps. The agent interleaves reasoning, action selection, tool interaction, and observation updates, and at each step decides whether to continue or terminate with a final answer or recommended action.
 
This loop matches many smart grid tasks, where the system cannot answer from internal knowledge alone and must call a solver, retrieve external information, and check intermediate results before reporting. Consider $N{-}1$ contingency screening, where the agent is asked whether the system can survive the loss of a specified transmission line. It first retrieves the current network topology and operating point, then invokes a power flow solver under the post-contingency configuration, inspects the returned bus voltages and line flows, and determines whether any operating limit is violated. If needed, it calls an OPF routine to evaluate corrective actions. Each step depends on the previous observation, and the number of steps is not known in advance.
 
Practical agentic systems extend the basic reasoning--action--observation loop with memory, which preserves prior context, intermediate observations, and learned preferences across steps and sessions, keeping the agent from repeating work or losing information between solver calls. For safety-relevant tasks, a verification step is often added before the final answer to confirm that reported values match the trusted tool output and that recommended actions satisfy task-level constraints. The remaining subsections develop the loop's components in turn: actions through API calls and tool use, common architectures, the role of memory, and the frameworks for building such systems.

\subsection{Action: API Calls and Tool Use}
\label{sec:action}
\begin{table}[tb]
\centering
\scriptsize
\caption{Common tool families used in agentic power system workflows}
\label{tab:common_ps_tools}
\renewcommand{\arraystretch}{1.15}
\setlength{\tabcolsep}{4pt}
\begin{tabularx}{\textwidth}{P{1.5cm} P{4.3cm} Y  P{3.7cm}}
\toprule
\textbf{Tool family} & \textbf{Example tools} & \textbf{Typical role in the workflow} & \textbf{Typical applications} \\
\midrule

Power system simulation and network analysis
& MATPOWER, Pandapower, PowerWorld, ANDES
& Load standard test cases or user-defined networks, run power flow simulations, and return critical physical quantities (voltage, angle, real and reactive power). Some tools also support dynamic simulation, OPF, and/or planning-oriented studies.
& Power flow analysis, contingency screening, GridDebug, ED/OPF studies, planning studies \\
\midrule

Optimization modeling and scheduling
& JuMP, CVXPY, Gurobi
& Formulate and solve constrained optimization problems such as economic dispatch, demand response, EV charging, and distributed energy resource sizing. These tools connect high-level model definitions to external solvers.
& Economic dispatch, residential demand response, EV charging scheduling, solar and storage sizing \\
\midrule

External data and retrieval interfaces
& {ISO/RTO market portals, utility tariff databases, regulatory and standards documents (e.g., FERC/NERC manuals and grid codes), weather and irradiance APIs, case repositories}
& Retrieve prices, emissions signals, irradiance, asset metadata, operating manuals, and other contextual information needed before optimization or explanation.
& Demand response, solar investment support, operator assistance, report generation \\
\midrule
Verification and validation
& Custom check scripts, feasibility and constraint validation, solver status parsers, unit/range checks, cross-tool consistency checks, task-specific validation routines
& Apply generic sanity checks (solver status, feasibility, units) and task-relevant validation (constraints, completeness, action criteria); typically task-specific routines.
& Applies across solver-grounded smart grid workflows  \\
\midrule
Visualization and reporting
& Python visualization libraries, dashboard modules, report generators
& Convert verified outputs into plots, tables, violation summaries, schedule views, and user-facing explanations.
& Power flow visualization, EV schedule reporting, solar scenario comparison, demand response summaries \\
\bottomrule
\end{tabularx}
\end{table}
Tool use is central to solver-grounded smart grid workflows~\cite{yao2023react,schick2023toolformer}. The principle is that whenever a task requires real-time data, exact computation, or external documents, the agent should call a tool rather than infer the answer from text patterns. In power systems, such tools include data APIs, power flow simulators, optimization solvers, and verification routines. The LLM selects the tool, formulates its input, interprets the output, and reports the result faithfully, while the tool provides the evidence. A practical tool interface makes this contract explicit through input and output schemas, unit conventions, and failure messages, which reduce common errors such as loading the wrong case or reporting a wrong unit.
 
It helps to distinguish \textit{read tools} from \textit{write tools}. Read tools retrieve or compute information without changing the environment, such as fetching weather forecasts, market prices, or event logs. Write tools modify it, for example by sending a control command, updating a schedule, or submitting a market bid. Because write tools create operational side effects, they should sit behind safeguards such as simulation, rule checking, approval workflows, or human-in-the-loop confirmation, which let agents support analysis, planning, and decision preparation while limiting unsafe or unauthorized actions. Tool use in smart grid agents is thus best understood as a task-specific grounding mechanism, not a generic capability: forecasting agents rely on data APIs and time-series models, scheduling agents on optimization solvers and feasibility checkers, and grid-analysis agents on power flow simulators, diagnostic parsers, and visualization modules. Table~\ref{tab:common_ps_tools} summarizes representative tool families and their roles.

\subsection{Common Agent Architectures}
\label{sec:architectures}
Agentic systems share a common workflow: acquire context, decide an action, call a tool, observe the result, update the plan. They differ in how they structure it. 
The choice of architecture depends on whether the task requires sequential evidence gathering, structured planning, hierarchical decomposition, error correction, or coordination across roles. Figure~\ref{fig:agent_architectures} summarizes the representative architectures along with typical roles in smart grids tasks.
\begin{figure}[tb]
    \centering
    \includegraphics[width=0.95\linewidth]{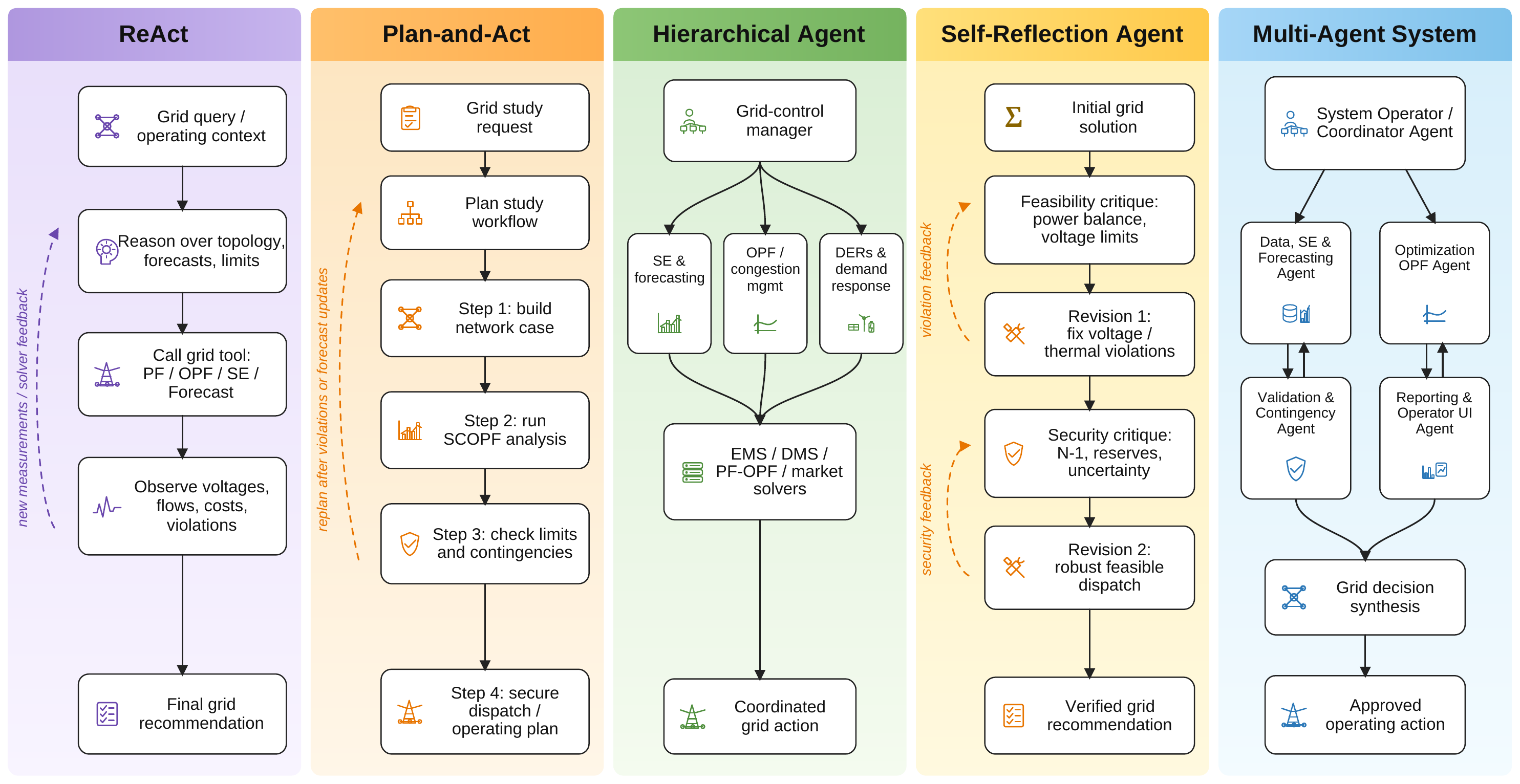}
    \caption{Comparison of representative agent architectures for smart-grid workflows. ReAct interleaves reasoning, action, and observation in a feedback loop. Plan-and-Act separates planning from stepwise execution. Hierarchical agents split the task into subtasks under a higher-level controller. Self-reflection agents revise drafts before the final answer. Multi-agent systems distribute the workflow across specialized agents.}
    \label{fig:agent_architectures}
\end{figure}

\textit{ReAct (Reasoning and Acting)} interleaves reasoning with tool actions in a tight feedback loop~\cite{yao2023react}. At each step the agent produces a short \emph{thought}, executes an \emph{action} such as calling a tool, and reads back the \emph{observation}. The observation feeds the next thought, and the loop ends when the agent judges the task complete. Keeping thought, action, and observation separate makes the trace inspectable rather than one opaque output. ReAct fits workflows with a small action space and local decisions, such as iterative diagnosis or short corrective-action loops (e.g., GridDebugAgent, Section~\ref{sec:griddebug}). Its limitation is the absence of a global plan. On long horizons the loop can drift or repeat steps, and reasoning at every step raises token cost.

{\textit{Plan-and-Act}} separates planning from execution~\cite{erdogan2025planact}. The agent first generates an explicit \emph{plan}, a structured list of steps with declared inputs, outputs, and dependencies, then executes the steps in order. Reasoning is invoked again only on failure, such as a solver not converging, at which point the agent revises the remaining plan and resumes. The separation suits workflows with clear stage boundaries, where each stage has a different cost profile. In smart grids, planning studies such as data center siting or transmission expansion planning fit naturally. The agent plans the full sequence of stages, then executes each in turn, replanning only if a stage fails. Its limitation is that the run is only as good as the upfront plan. When the setting is hard to anticipate, an inaccurate plan forces repeated replanning.
 
\textit{Hierarchical agents} organize decision making into levels~\cite{ichter2023saycan,liu2023hla}. A higher-level \emph{controller} sets the goal and decomposes it into subtasks, while lower-level \emph{workers} each handle one subtask with their own tools. The controller aggregates worker outputs and decides whether to spawn more subtasks, revise the decomposition, or terminate. The pattern fits workflows that split into many independent subtasks whose results combine into one answer. In smart grids, $N{-}1$ contingency screening is a natural example. The controller dispatches one worker per outage to run the post-contingency power flow, aggregates the results into a ranked list of the most dangerous contingencies, and may spawn corrective-action workers for the critical cases. Its limitation is decomposition overhead. The controller must coordinate and merge worker outputs, and a poor decomposition degrades the answer.
 
\textit{Self-reflection agents} insert a self-critique step between the draft answer and the final output. After an initial response, the agent invokes itself as a critic, examines the draft for errors or constraint violations, and either accepts it or revises. Reflexion stores feedback across trajectories to guide future attempts~\cite{shinn2023reflexion}. Self-Refine critiques and refines the current output in place~\cite{madaan2023selfrefine}. In smart grids, self-reflection adds an internal check for unsupported claims or inconsistencies between the answer and the tool evidence. This suits operator-facing reporting. An agent drafting an OPF or contingency summary can self-critique to catch a number that does not trace to the solver output, and revise before finalizing. Because the critic is itself an LLM, self-reflection alone is not sufficient and should be paired with a verification layer for solver-grounded correctness.

\textit{Multi-agent systems} partition the workflow across specialized agents with defined roles and interfaces~\cite{wu2023autogen}. Each agent has its own prompt, tool set, and reasoning loop, often orchestrated by a coordinator that dispatches tasks and aggregates results. For example, the data center siting workflow above can be re-cast in multi-agent form, with separate data, optimization, validation, and reporting agents each owning a stage. This adds auditable per-stage interfaces at the cost of coordination overhead, and pays off when stages demand independent expertise. Otherwise a single-agent architecture is simpler and more robust.

Are these architectures themselves trainable? They are orchestration patterns, not models: each runs on an off-the-shelf LLM through prompting and control flow, which is why the case studies in Section~\ref{sec:casestudies} use frontier models without fine-tuning. What can be trained is the underlying LLM, which SFT adapts for tool calling and structured output and which verifiable-reward fine-tuning can specialize into smaller tool-calling agents~\cite{lambert2024tulu3}. Across all these architectures, the key requirement for power systems stays solver-grounded correctness: numerical results should originate from trusted tools and pass verification before they are reported.

\subsection{Memory Management}
\label{sec:memory}
Memory lets an agent preserve state across steps: what has been tried, which tools were called, what observations returned, and whether outputs passed verification. It should be treated as a structured system component, not just a longer conversation history~\cite{shinn2023reflexion,park2023generative,packer2023memgpt}. It includes \emph{short-term memory}, holding state within a single trajectory, and \emph{long-term memory}, which carries across trajectories and holds episodic and semantic memories.
 
\emph{Short-term memory} holds the working state within a single task: a scratchpad recording the current goal, steps taken, tool outputs received, and intermediate conclusions. Each reasoning step reads from it and writes back new information. Because it persists only for one trajectory, it is most exposed to context-window pressure: when the working state grows too large, items can be dropped or paraphrased, silently corrupting the reasoning~\cite{packer2023memgpt}. In smart grid workflows, it keeps operational details such as the active network case and the latest solver output accessible across steps within a task.

\emph{Long-term memory} carries state across tasks, typically as an external store~\cite{packer2023memgpt}, with two subtypes. \emph{Episodic memory} is the agent's logbook of past runs, storing the task, tool calls, observations, failures, and the final verified outcome of each~\cite{shinn2023reflexion}. \emph{Semantic memory} stores patterns and rules distilled from many runs: reusable guidance (``always verify solver status before reporting''), report templates, common causes of non-convergence, and standard corrective actions~\cite{park2023generative}. It gives consistency across tasks, but a faulty rule generalized from one episode propagates to every later run, so updates must be cautious. Verification status is an important filter: values from runs that failed verification should never be promoted to semantic memory as ground truth.

\subsection{Frameworks and Tools for Building Agents}
\label{sec:frameworks}
Frameworks for agentic AI provide the engineering layer for defining agents, registering tools, managing memory, and deploying workflows. We group them by what they orchestrate, since the unit of orchestration most shapes how a smart grid workflow is built. Three classes differ in this unit: a single agent's loop, a team of agents, or an explicit workflow graph, all relying on a connectivity layer of interoperability protocols. Because the ecosystem changes quickly, we describe each class by its durable engineering role and name representative tools only as illustrations.

Agent SDKs and middleware orchestrate a single agent's loop, with lightweight abstractions for tool use and logging (OpenAI Agents SDK~\cite{openai2025agentssdk}, Semantic Kernel~\cite{microsoft2025semantickernel}, Strands Agents~\cite{aws2025strands}). They are the fastest to adopt but leave control flow implicit. Multi-agent frameworks orchestrate a team of agents, with abstractions for inter-agent messaging and role assignment (AutoGen~\cite{wu2023autogen}, CrewAI~\cite{crewai2025}). They suit problems that decompose into distinct roles. Workflow orchestrators run an explicit graph of steps, with declared transitions, retries, and checkpoints (LangGraph~\cite{langgraph2025}, Microsoft Agent Framework~\cite{microsoft2025agentframework}). They give the most control at the cost of more setup, and their explicit steps make a natural place to insert a verification gate. The three classes are not mutually exclusive. Interoperability protocols form the connectivity layer, standardizing agent-to-tool and agent-to-agent connections (MCP~\cite{anthropic2024mcp}, A2A~\cite{google2025a2a}). They orchestrate and verify nothing themselves but provide the substrate the framework classes use when external tools or remote agents enter the loop.
 
In practice, power-system agents have favored lightweight custom orchestration over heavyweight frameworks. Most are bespoke single-agent loops with direct function calling, as in GridDebugAgent and PFAgent, or multi-agent designs that separate planning from validation, as in GridMind~\cite{jin2025gridmind} and Grid-Agent~\cite{zhang2025gridagent}. At the connectivity layer, MCP is emerging as the common substrate, adopted by X-GridAgent~\cite{wen2025xgridagent} and advocated in the PowerAgent roadmap~\cite{zhang2025poweragent}. Two patterns stand out. Framework choice is rarely standardized or reported across studies, which complicates reproducibility. Verification is seldom part of the framework layer, so it must be added explicitly.

The unit of orchestration guides which framework class to choose:
\begin{itemize}
\setlength{\itemsep}{0pt}\setlength{\parskip}{0pt}
\item \textbf{Agent SDKs} for single-agent assistants, such as operator copilots, report generation, and the conversational PFAgent and GridDebugAgent of our case studies.
\item \textbf{Multi-agent frameworks} when the workflow decomposes into distinct roles, as in GridMind and Grid-Agent, which separate planning from validation.
\item \textbf{Workflow orchestrators} when a task has clear stage boundaries and steps must be verified before execution, such as multi-stage planning studies or $N{-}1$ contingency screening.
\item \textbf{Interoperability protocols} whenever external tools or remote agents must connect, as in the MCP-based interface of X-GridAgent, independent of which framework orchestrates the task.
\end{itemize}
In every case the framework supplies only the engineering layer: solver-grounded correctness still rests on the verification gate and trusted tools of Section~\ref{sec:workflow}, not the framework, and Section~\ref{sec:evaluation} discusses evaluation and verification.

\section{Evaluation Protocol}
\label{sec:evaluation}
Evaluation of LLM-based and agentic systems for smart grids must match the solver-grounded workflow of Section~\ref{sec:workflow}. A system that produces fluent but unverified numbers can score well on standard language metrics while still issuing physically infeasible decisions, and existing evaluation efforts were not designed for this setting. General agent benchmarks measure end-task success and tool-use competence but stay domain-agnostic and do not tie a tool's output to physical feasibility~\cite{yao2024taubench}. RAG metrics assess faithfulness to retrieved text rather than to a trusted solver~\cite{es2024ragas}. Selective-prediction criteria formalize abstention for classification rather than for verification-gated numerical reporting~\cite{geifman2017selective}. Recent surveys confirm that agent evaluation remains fragmented, and that agent-level hallucinations propagate across tool calls and reasoning steps in ways single-turn metrics miss~\cite{mohammadi2025agenteval,lin2025agenthallucinations}. To our knowledge, no existing framework couples these concerns into one protocol linking tool-output correctness to grid feasibility, the gap the protocol below addresses.

We organize the evaluation around four metric groups that follow from the solver-grounded rule (Section~\ref{sec:solver-grounded}). 
The rule reports a number only when a trusted tool computes it and verification passes. A numerical claim can therefore fail at three successive points on its path from computation to report. It can be inaccurate on the objective, infeasible at the tool boundary, or misreported in the final text. Producing a verified answer also consumes resources such as tokens, solver compute, and time. These four concerns define the groups. (i)~\emph{Task utility} is accuracy on the primary modeling or optimization objective. (ii)~\emph{Solver-grounded correctness} is whether the solution respects physical and operational constraints at the tool boundary. (iii)~\emph{Faithfulness and safe failure} is whether the natural-language report transmits the verified output without distortion and abstains when verification fails. (iv)~\emph{Cost and latency} is the resources required to produce a verified answer.

Each group exposes a failure mode invisible to the others. A low-error forecast can still violate a constraint, a feasible solver result can still be misreported, and a correct, faithful answer can still be too slow or costly to deploy. Groups (i) to (iii) exhaust the points at which a numerical claim can go wrong under the rule, while (iv) is the one concern orthogonal to correctness that gates deployment. Task utility alone, the metric most often reported in prior capability studies~\cite{huang2024foundation,majumder2024electric}, hides exactly the constraint violations and report-level hallucinations that groups (ii) and (iii) expose.

\subsection{Task utility}
\label{subsec:eval-utility}
Task utility measures how good an answer is on the problem's own terms, independent of how it was produced. It is the score we would assign if no LLM or agent were involved, reported identically for the LLM-only and solver-grounded variants so the two are directly comparable. For regression and forecasting tasks, we report a prediction-error metric against ground truth, such as mean absolute error (MAE), root-mean-square error (RMSE), or mean absolute percentage error. For classification and detection subtasks, we report an accuracy- or overlap-based metric such as F1 score. For constrained optimization, we report the achieved objective as an optimality gap relative to the optimal solver solution, paired with a task-completion metric such as the fraction of demand served. For diagnosis and remediation, we report a success metric such as repair rate with a partial-progress metric such as violation reduction. These families cover the most common smart grid tasks, and a task outside them adopts the native metric for its objective. Task utility is necessary but not sufficient. A standalone LLM can minimize MAE by interpolating recent history while still violating physical constraints, which is why the remaining three groups are needed.
Some tasks are multi-objective, so a single scalar does not suffice. EV scheduling trades cost against unmet energy, and power flow reports voltage, flow, and balance errors. For these, task utility is a small vector of complementary metrics rather than one number.

\subsection{Solver-grounded correctness}
\label{subsec:eval-correctness}
Solver-grounded correctness measures whether the tool-derived solution is valid on its own terms: whether it respects the system's physical and operational constraints, and whether the solver's outcome (optimal, infeasible, or non-converged) is correctly determined. We read these off the task-level verification $\mathcal{V}(x,c,z,y)$ of Section~\ref{sec:solver-grounded}. Writing $\mathcal{V}(x,c,z,y)=(b,\rho)\in\{\text{pass},\text{fail}\}\times\mathbb{R}^m$ from~\eqref{eq:verification}, $b$ is the verification status and $\rho$ collects the residuals of the $m$ conditions. A check passes when its residual lies within tolerance, and $b=\text{pass}$ only when all pass. We partition the conditions into \emph{constraint-satisfaction checks} $\mathcal{C}$, testing physical and operational limits, and \emph{faithfulness checks} $\mathcal{F}$, testing whether the reported numbers match the tool output, with $\mathcal{C}\cup\mathcal{F}=\{1,\dots,m\}$. This subsection quantifies $\mathcal{C}$ and the reported solver status. Faithfulness (Section~\ref{subsec:eval-faithfulness}) quantifies $\mathcal{F}$. For run $i$, write $v_i=(b_i,\rho_i)=\mathcal{V}(x_i,c_i,z_i,y_i)$ and let $\mathbf{1}[\mathrm{feas}_i]=1$ indicate that run $i$ passes every check in $\mathcal{C}$. Then $b_i=\text{pass}$ iff $\mathrm{feas}_i=1$ and run $i$ also passes every check in $\mathcal{F}$. The following metrics quantify correctness at the tool boundary:
\begin{itemize}
    \item \textbf{Feasibility rate:}
    \begin{equation}
        \mathrm{FR} \;=\; \frac{1}{N}\sum_{i=1}^{N} \mathbf{1}\!\left[\mathrm{feas}_i\right],
        \label{eq:feasibility_rate}
    \end{equation}
    the fraction of runs whose tool-derived solution passes all constraint-satisfaction checks.
    \item \textbf{Maximum constraint violation:} $\max_{j\in\mathcal{C}} \, [\rho_i]_j$, the worst residual or limit violation across the constraint-satisfaction checks, reported even when $\mathrm{FR}=1$ to characterize operational margins.
    \item \textbf{Solver status accuracy:} whether the reported solver status is correct, classified among $\{\text{optimal},\allowbreak\ \text{infeasible},\allowbreak\ \text{non-converged}\}$ and evaluated as a confusion matrix over the scenario set. Misreporting a non-converged case as optimal is the most damaging error class and should be tracked separately.
    \item \textbf{Trial reliability ($\mathrm{pass}^k$):} for stochastic agent pipelines, the fraction of scenarios where all $k$ independent trials are feasible ($\mathbf{1}[\mathrm{feas}]=1$ on every trial)~\cite{yao2024taubench}. Even at a high single-trial feasibility rate, a low $\mathrm{pass}^k$ shows the agent cannot be trusted to act consistently across repeated invocations, a critical requirement for operational deployment.
\end{itemize}
Solver-grounded correctness matters most in scheduling and power flow workflows, where an infeasible solution can pass undetected into downstream control decisions.

\subsection{Faithfulness and safe failure}
\label{subsec:eval-faithfulness}
A solver-grounded system must report what the tool actually produced (the faithfulness checks $\mathcal{F}$ of Section~\ref{subsec:eval-correctness}), and must abstain when verification fails. These two requirements correspond to established ideas from RAG evaluation~\cite{es2024ragas} and selective prediction~\cite{geifman2017selective}, adapted here to the tool-use setting.
\begin{itemize}
    \item \textbf{Faithfulness:} this metric instantiates the checks in $\mathcal{F}$, one per reported number. Let $R = \{r_1, \dots, r_K\}$ be the numerical values appearing in the system's natural-language report and $\{z_1, \dots, z_K\}$ the corresponding verified tool values. We define
    \begin{equation}
        \mathrm{Faith} \;=\; \frac{1}{K}\sum_{k=1}^{K} \mathbf{1}\!\left[|r_k - z_k| \le \epsilon_k\right],
        \label{eq:faithfulness}
    \end{equation}
    where $\epsilon_k$ is a quantity-specific numerical tolerance (e.g., $10^{-3}$~p.u.\ for voltage magnitudes, $1\%$ for branch flows, $\$0.01$ for costs). $\mathrm{Faith}=1$ means every reported number is traceable to a verified tool output. Values below~1 indicate hallucinated, rounded, or cross-referenced numbers.
    \item \textbf{Safe-failure rate:} among runs whose verification fails ($b=\text{fail}$ in~\eqref{eq:verification}), the fraction in which the system explicitly abstains from reporting numerical results, instead stating the failure and its cause,
    \begin{equation}
        \mathrm{SFR} \;=\; \Pr\!\left[\text{abstain} \,\big|\, \mathcal{V}(x,c,z,y)=\text{fail}\right].
        \label{eq:safe_failure}
    \end{equation}
\end{itemize}
Together with the correctness metrics of Section~\ref{subsec:eval-correctness}, faithfulness and safe failure operationalize the design rule of Section~\ref{sec:solver-grounded}: a number reaches the report only if it comes from a feasible, verified tool output and is transcribed without distortion. Otherwise the system flags the failure. Unlike single-turn LLM hallucinations, agent hallucinations propagate across tool calls, memory updates, and reasoning steps, creating failure modes such as tool-output misreading, cross-trial contamination, and intermediate-step fabrication that have only recently been taxonomized~\cite{lin2025agenthallucinations}. Verification-gated reporting addresses the worst of these by construction: the agent cannot report a number no trusted tool produced. In our case studies these properties are audited, not asserted. For the three solver-backed tasks, every reported metric is recomputed by re-running the trusted tool on the system the agent produced, and the same checker scores both baseline and agent, so a confident but wrong claim cannot inflate the score. In the no-solver forecasting case, the audit reduces to output-level checks (schema validity and range).

\subsection{Cost and latency}
\label{subsec:eval-cost}
Cost and latency capture the resources required to produce a verified answer, motivated by two concerns. First, the per-request footprint: at scale, each request consumes priced LLM tokens, tool and solver compute, and tool invocations, so a correct but expensive system may be impractical to run continuously. Second, responsiveness: in time-sensitive settings such as real-time dispatch or contingency screening, the operator needs a verified answer within a useful time window.
\begin{itemize}
    \item \textbf{Tool calls per request} ($n_{\text{tool}}$): the number of external tool invocations, such as solver, simulator, retriever, or checker calls, that the system issues while answering one request, counting every invocation in the trajectory, including retries triggered by failed verifications.
    \item \textbf{Token cost per request}: the total number of LLM tokens consumed while answering one request, summed over all model calls in the trajectory, and reported with prompt, completion, and retrieval context tokens listed separately so that long context and RAG systems can be compared on a common basis.
    \item \textbf{Time to verified answer:} the wall-clock time from the user request to a verified report, summing generation, tool-execution, and verification time over the $n_{\text{tool}}$ iterations of the agent loop,
   \begin{equation}
        \mathrm{TTV} \;=\; \sum_{k=1}^{n_{\text{tool}}} \!\left(\tau_{\text{llm},k} + \tau_{\text{tool},k} + \tau_{\text{verify},k}\right),
        \label{eq:ttv}
    \end{equation}
\end{itemize}

\section{Case Studies}
\label{sec:casestudies}

This section presents four case studies that ground the preceding framework in concrete implementations. Each study describes the problem setup, the system design, and the evaluation results.
\subsection{LLM for Wind Power Forecasting}
\paragraph{\textbf{Problem Description}}
  Wind power forecasting predicts turbine active power output over a future horizon from historical wind power measurements and meteorological conditions. This case study evaluates whether LLMs can forecast time series by learning wind-to-power relationships from historical context. We use the SDWPF dataset \cite{zhou2024sdwpf}: 134 wind turbines sampled at 10-minute intervals over 245 days (May--December 2020). Each observation includes wind speed (Wspd), wind direction (Wdir), external temperature (Etmp), and active power output (Patv).

The forecasting input is a 14-day historical context window, $\mathbf{X}_{t-W:t} = \{\mathrm{Wspd}, \mathrm{Wdir}, \mathrm{Etmp}, \mathrm{Patv}\}_{t-W:t}$, plus, when available, a 48-hour meteorological forecast $\mathbf{F}_{t+1:t+H}$ based on ERA5 100m wind speed~\cite{ERA5}. The task is to predict the next 48 hours of active power output. Here $W = 2{,}016$ steps is 14 days at 10-minute resolution and $H = 288$ steps is 48 hours. The model must return exactly 288 numerical values in JSON format: \texttt{\{"forecast": [f1, f2, ..., f288]\}}.

  \paragraph*{\textbf{Strategy}}
  We compare three prompting approaches for LLM-based wind forecasting, where the LLM is the prediction engine rather than an orchestrator of external solvers. \textbf{(1) Baseline (Naive Prompt):} sends raw 14-day wind power data (2,016 timesteps $\times$ 4 features) directly to the LLM with minimal instruction: ``Predict the next 48 hours of active power. Return 288 numerical values.'' No physical constraints, external weather data, or output-schema enforcement. \textbf{(2) Advanced (Advanced Prompt):} augments the naive prompt with injected physical priors (power $\propto$ wind speed$^3$ at low speeds, saturation near 1500\,kW, near-zero output below 3\,m/s) and a strict JSON output schema enforced with retry logic (up to 3 attempts), but without time-series binning or external meteorological context. 
\textbf{(3) APBF (Advanced Prompt + Binning + Forecast):} builds on the advanced prompt with \textit{Binning}  (Discretizes wind speed and power into 16 ordinal levels (0-15) to reduce token count by $\sim$50\%.) and \textit{forecast with meteorological features} (Retrieves 48-hour wind speed forecasts (100m altitude) from Open-Meteo ERA5 API, interpolates to 10-minute resolution, and appends to the prompt as external meteorological contexts for the forecast). Relating these to Section~\ref{sec:prompting}, the naive prompt is a plain instruction, the advanced prompt applies structured prompting (Section~\ref{sec:structured-prompting}) through an output schema and injected physical priors, and APBF augments the prompt with external context retrieved from a data API rather than a document store.
Fig.~\ref{fig:wind_trajectory} shows a representative APBF execution trace:
  \begin{figure}[tb]
  \centering
  \begin{minipage}{0.94\linewidth}
  \begin{promptfigure}{Wind Forecasting Trajectory (APBF Configuration, Turbine 1, Day
  1)}
  \scriptsize
  {\ttfamily\scriptsize
  \rolehl{User:} Predict Day 15--16 active power for Turbine 1
  (Base\_Day=1).
  \reasonhl{System:} Binning 14-day wind power history into 16 levels...\contexthl{Tool Call (Open-Meteo):} get\_openmeteo\_wind\_data(lat=40.6306, lon=96.9498, start="2020-05-15", end="2020-05-16")\par
  \retrievedhl{Tool Output:} [wind\_speed\_100m: 4.2, 5.1, 7.8, 8.3, ...] (48h
  hourly)\par
  \reasonhl{System:} Interpolating to 288 10-min steps. Constructing prompt: 
  \neutralprompt{\textit{Context}: You are an expert wind turbine power forecasting model. } \neutralprompt{\textit{Input 1}: 14 days of historical SCADA (Day 1--14, 144 rows/day):... [2016 rows total]}
  \neutralprompt{Input 2: 48-Hour Meteorological Forecast (100m wind, m/s):[4.2, 4.3, 4.5, ..., 8.1] (288 values)}
  \neutralprompt{Task: Predict Patv (kW) for NEXT 48 HOURS (288 timesteps).}
  \neutralprompt{Instructions: Power $\propto$ wind speed$^3$ at low speeds, Power saturates near 1500 kW, Power near zero below 3 m/s, Do NOT copy the last day, Clip values to [0, 1500]}
  \neutralprompt{OUTPUT FORMAT (CRITICAL): \{"forecast": [f1, f2, ..., f288]\} Exactly 288 numbers. No text outside JSON. No markdown.}
  \outputhl{LLM Output:} \texttt{\{"forecast": [480.2, 510.5, 545.8, ..., 892.1]\}}\\
  \reasonhl{Validation:} Length = 288 \checkmark; All values in [0,1500]
  \checkmark\par
  \outputhl{Metrics (vs. ground truth):} MAE = 189.8 kW, RMSE = 292.3 kW
  }
  \end{promptfigure}
  \end{minipage}
  \caption{Wind forecasting trajectory for the proposed APBF method (Advanced Prompt + Binning + Forecast): the agent bins history, retrieves a weather forecast, builds a physics-informed prompt, validates the JSON schema, and scores against ground truth.}
  \label{fig:wind_trajectory}
  \end{figure}

  \paragraph*{\textbf{Results}}
We compare three prompting strategies (Naive, Advanced, APBF) across three frontier LLM families (Gemini~3 Flash, GPT-5.4, Claude Sonnet~4.6) and a GRU baseline, the baseline from the KDD Cup 2022 Wind Power Forecasting Challenge on the SDWPF dataset~\cite{zhou2024sdwpf}. All methods are evaluated over 3, 6, and 48~hour horizons on held-out turbines. Table~\ref{tab:wind-complete_results} reports MAE and RMSE for each model-strategy-horizon combination with an overall score (mean of MAE and RMSE). Fig.~S1 in the Supplementary Material adds a representative time series, the 48-hour LLM forecast against ground-truth active power for a single turbine.
 
Three patterns emerge. \emph{First}, the GRU baseline stays competitive across all horizons, so LLM-based models still have room to improve on time-series prediction. \emph{Second}, APBF consistently improves long-horizon prediction (e.g., 48\,h) over naive prompting, showing that physical constraints, proper tokenization, and external weather context can materially reduce error. \emph{Third}, no single model-strategy combination dominates across horizons: Gemini~3 + Naive is best at 3~h (147.21), Claude Sonnet + Advanced at 6~h (244.49), and Gemini~3 + APBF at 48~h.

This sensitivity to prompting strategy and model choice matches the broader pattern across the case studies. Advanced prompting shapes LLM behavior but does not enforce physical correctness or guarantee gains, and no single model dominates, so LLM selection should rest on empirical benchmarking of the target task. The LLM approach still offers practical advantages over conventional ML models like the GRU: it ingests heterogeneous context (historical turbine data, weather forecasts, natural-language physical priors) in a single inference call without per-feature engineering, and adapts to a new wind farm by editing the prompt rather than retraining.

\begin{table}[!ht]
\centering
\caption{Overall forecasting performance for all Model-Strategy-Horizon combinations. For each model, the lowest Overall per horizon is in \textbf{bold}.}
\label{tab:wind-complete_results}
\resizebox{\textwidth}{!}{%
\begin{tabular}{llccccccccc}
\toprule
\multirow{2}{*}{\textbf{Model}} & \multirow{2}{*}{\textbf{Strategy}} 
& \multicolumn{3}{c}{\textbf{3h}} 
& \multicolumn{3}{c}{\textbf{6h}} 
& \multicolumn{3}{c}{\textbf{48h}} \\
\cmidrule(lr){3-5} \cmidrule(lr){6-8} \cmidrule(lr){9-11}
& & \textbf{MAE (kW)} & \textbf{RMSE (kW)} & \textbf{Overall}
  & \textbf{MAE (kW)} & \textbf{RMSE (kW)} & \textbf{Overall}
  & \textbf{MAE (kW)} & \textbf{RMSE (kW)} & \textbf{Overall} \\
\midrule

\multirow{3}{*}{Gemini 3 Flash}
& Naive    & 110.56 & 183.86 & \bf{147.21} & 383.72 & 430.22 & 406.97 & 1030.67 & 1114.29 & 1072.48 \\
& Advanced & 516.72 & 532.90 & 524.81 & 361.00 & 403.75 & \bf{382.38} & 366.22 & 439.44 & 402.83 \\
& APBF     & 422.11 & 455.56 & 438.83 & 413.88 & 455.67 & 434.78 & 189.79 & 292.34 & \bf{241.07} \\
\midrule

\multirow{3}{*}{GPT 5.4}
& Naive    & 202.22 & 259.66 & 230.94 & 658.91 & 821.68 & 740.30 & 676.47 & 776.01 & 726.24 \\
& Advanced & 171.20 & 216.03 & 193.61 & 491.01 & 610.43 & \bf{550.72} & 648.14 & 744.14 & 696.14 \\
& APBF     & 144.19 & 192.64 & \bf{168.41} & 557.52 & 716.23 & 636.87 & 445.58 & 598.04 & \bf{521.81} \\
\midrule

\multirow{3}{*}{Claude Sonnet 4.6}
& Naive    & 335.34 & 397.44 & 366.39 & 267.47 & 348.58 & 308.03 & 446.13 & 615.07 & 530.60 \\
& Advanced & 141.20 & 192.82 & \bf{167.01} & 201.08 & 287.90 & \bf{244.49} & 580.67 & 740.53 & 660.60 \\
& APBF     & 1101.93 & 1108.02 & 1104.98 & 530.04 & 586.12 & 558.08 & 316.74 & 390.96 & \bf{353.85} \\
\midrule

GRU Baseline & - 
& 144.82 & 162.68 & 153.75 & 191.54 & 216.08 & 203.81 & 351.34 & 280.28 & 315.82 \\
\bottomrule
\end{tabular}}
\end{table}

\subsection{EV Scheduling Agent}

\paragraph*{\textbf{Problem Description}}
EV charging scheduling allocates power to concurrent sessions so each vehicle receives as much of its requested energy as possible before departure, while respecting per-vehicle charging power limits and a site-level capacity cap. Formally, consider scheduling over a day $\mathcal{T} = \{0,\ldots,95\}$ (96 steps of 15\,min, 24 hours total). Each charging session $i \in \mathcal{I}$ has arrival $a_i$, departure $d_i$, requested energy $E_i^{\mathrm{req}}$ [kWh], and per-session power cap $\bar{p}_i$ [kW] (dependent on the vehicle type). The site has aggregate cap $\bar{P}_t$ [kW] and TOU price $c_t$ [\$/kWh]:
 \begin{align}
 \min_{\{p_{i,t}\},\{u_i\}} \quad &
     \Delta t \sum_t c_t \sum_i p_{i,t}
     + \lambda \sum_i u_i \label{eq:ev_obj}\\
 \text{s.t.}\quad
 & p_{i,t} = 0, \quad t \notin [a_i, d_i) \label{eq:ev_avail}\\
 & 0 \le p_{i,t} \le \bar{p}_i \label{eq:ev_power}\\
 & \textstyle\sum_i p_{i,t} \le \bar{P}_t \label{eq:ev_cap}\\
 & \Delta t \textstyle\sum_{t=a_i}^{d_i-1} p_{i,t} + u_i
     = E_i^{\mathrm{req}}, \quad u_i \ge 0. \label{eq:ev_energy}
 \end{align} 
 Here $p_{i,t}$ is the charging power of session $i$ in slot $t$, $\Delta t$ is the slot duration, and $\lambda$ [\$/kWh] is the penalty weight. The slack $u_i \ge 0$ permits unmet energy but is penalized in the objective function.

 \paragraph*{\textbf{Strategy}}
We compare three approaches on identical session data, all using the same CVXPY solver for~(\ref{eq:ev_obj})--(\ref{eq:ev_energy}) with $\Delta t = 0.25$\,h and $\lambda = 10^6$\,\$/kWh (a globally optimal linear program)~\cite{diamond2016cvxpy}. (1) \textit{Numerical solver}: CVXPY called directly on structured session data, the gold standard. (2) \textit{LLM-only baseline}: GPT-4o or Sonnet~4.6 receives the problem as natural-language text and must output the schedule matrix directly, with no optimizer or tools. (3) \textit{EVAgent}: GPT-4o or Claude Sonnet~4.6 as the backbone LLM, with tool access to the CVXPY solver. 

EVAgent operates in three stages within a multi-turn loop. \textit{Parsing:} An LLM call receives the problem as natural-language text and extracts structured per-session parameters (arrival/departure step indices, energy, max power). Missing fields trigger a user clarification request. \textit{Optimize (CVXPY tool):} For scheduling requests, the LLM autonomously invokes \texttt{solve\_ev\_schedule} with the parsed parameters. The tool solves~(\ref{eq:ev_obj})--(\ref{eq:ev_energy}) and returns a JSON summary (\texttt{total\_cost\_usd}, \texttt{peak\_load\_kw}, \texttt{total\_unmet\_kwh}, \texttt{pct\_fully\_served}) with the full schedule matrix. \textit{Explain:} The LLM synthesizes the tool's JSON output into a natural-language summary. Every numerical claim traces to a specific tool return value, preventing metric hallucination.

This architecture enforces the solver-grounded rule: every reported number originates from the verified CVXPY tool, not from LLM inference. Fig.~\ref{fig:ev_trajectory} shows a complete {EVAgent} trajectory for a 7-EV natural-language request.

\begin{figure}[htb]
\centering
\begin{minipage}{0.94\linewidth}
\begin{promptfigure}{EVAgent Trajectory, 7-EV Free-Form Request (Caltech ACN-Data JPL Site)}
\scriptsize
{\ttfamily\scriptsize
\rolehl{User:} I have 7 EVs: EV1 arrives 6\,pm leaves 10\,pm needs 20\,kWh, EV2 arrives 7\,pm leaves 11\,pm needs 15\,kWh, EV3 arrives 8\,pm leaves midnight needs 10\,kWh, EV4 arrives 3\,pm leaves 4\,pm needs 3\,kWh, EV5 arrives 5\,pm leaves 7:30\,pm needs 3\,kWh, EV6 arrives noon leaves 6\,pm needs 30\,kWh, EV7 arrives 8\,am leaves 11\,am needs 6\,kWh.\par
\reasonhl{Agent (Parse):} Mapped 7 sessions to 15-min step indices. EV1: steps 72-87, EV2: steps 76-92, EV3: steps 80-95, EV4: steps 60-63, EV5: steps 68--74, EV6: steps 48-72, EV7: steps 32-44. Per session power caps loaded
\contexthl{Tool call:} \texttt{solve\_ev\_schedule(\{sessions: [7 entries], site\_cap\_kw: 50, tou\_off\_peak: 0.12, tou\_peak: 0.45\})}\par
\retrievedhl{Tool output:} \texttt{\{total\_cost\_usd: 18.96, peak\_load\_kw: 17.38, total\_unmet\_kwh: 0.0, pct\_fully\_served: 100.0\}}\par
\contexthl{Checker:} 0 hard constraint violations; 0 energy shortfall flags.\par
\outputhl{Agent (Explain):} ``The charging schedule serves all 7 sessions fully with no unmet energy. Total cost: \$18.96. Peak load: 17.38\,kW, well below the 50\,kW site cap. Sessions with earlier departure windows (EV4, EV7) are charged off-peak to minimize TOU cost.''}
\end{promptfigure}
\end{minipage}
\caption{An {EVAgent} trajectory for a 7-EV free-form request: the agent maps time expressions to step indices, calls \texttt{solve\_ev\_schedule}, passes the result to the checker, and explains it using only tool-JSON numbers. All 7 sessions are fully served; load peaks at 17.38\,kW, below the 50\,kW cap.}
\label{fig:ev_trajectory}
\end{figure}

\paragraph*{\textbf{Results}}

We compare three methods. 
\emph{Checker flags} counts all violations the shared checker returns across sessions and time steps, combining \emph{hard constraint violations} on~(\ref{eq:ev_avail})--(\ref{eq:ev_cap}) (charging outside session windows, per-charger or site-cap exceedances) and \emph{energy shortfall flags} where $u_i > 0$ (under-delivery forced by the site cap). The numerical solver and {EVAgent} incur only the latter on high-congestion days, while the LLM-only baseline also triggers hard violations. The evaluation uses 20 benchmark days from the Caltech ACN-Data JPL site~\cite{lee2019acndata}. TOU prices are \$0.12/kWh off-peak (steps 0-63, 84-95) and \$0.45/kWh peak (steps 64-83, i.e., 4-9\,pm).

\begin{table}[ht]
\centering
\caption{EVAgent vs.\ LLM-only and the CVXPY solver on 20 Caltech ACN-Data JPL days (mean of 5 runs). \textbf{Cost}: TOU cost; \textbf{Peak}: max site load (50\,kW cap); \textbf{Unmet}: undelivered energy ($\sum_i u_i$); \textbf{Served}: fully served sessions; \textbf{Flags}: checker violations (hard $+$ shortfall, $u_i>0$). \textbf{Bold}: solver-grounded feasible optimum; $^\dagger$lower LLM-only cost/peak are under-delivery artifacts.}
\label{tab:ev_results}
\scriptsize
\setlength{\tabcolsep}{2pt}
\renewcommand{\arraystretch}{1.12}
\begin{tabularx}{\linewidth}{@{}l l Y Y Y Y Y@{}}
\toprule
\textbf{Model} & \textbf{Method}
  & \textbf{Cost (\$)~$\downarrow$}
  & \textbf{Peak (kW)~$\downarrow$}
  & \textbf{Unmet (kWh)~$\downarrow$}
  & \textbf{Served (\%)~$\uparrow$}
  & \textbf{Flags~$\downarrow$} \\
\midrule
Numerical Solver & ---
  & \textbf{100.83}
  & \textbf{47.35}
  & \textbf{27.82}
  & \textbf{71.31}
  & \textbf{15.80} \\
\midrule
\multirow{2}{*}{GPT-4o}
  & LLM-Only
  & $95.29^{\dagger}$
  & $76.90^{\dagger}$
  & $209.47$
  & $36.36$
  & $121.62$ \\
  & {EVAgent}
  & \textbf{100.83}
  & \textbf{47.35}
  & \textbf{27.82}
  & \textbf{71.31}
  & \textbf{15.80} \\
\midrule
\multirow{2}{*}{Sonnet 4.6}
  & LLM-Only
  & $62.84^{\dagger}$
  & $28.07^{\dagger}$
  & $264.61$
  & $27.93$
  & $97.54$ \\
  & {EVAgent}
  & \textbf{100.83}
  & \textbf{47.35}
  & \textbf{27.82}
  & \textbf{71.31}
  & \textbf{15.80} \\
\bottomrule
\end{tabularx}
\end{table}

Table~\ref{tab:ev_results} shows that {EVAgent} reproduces the CVXPY solution on all five metrics across all 20 days on both GPT-4o and Sonnet 4.6 backends, with no failed or mis-parsed tool calls. Reproducing the solver is the expected outcome rather than a finding, since the agent calls CVXPY instead of optimizing itself. What the table measures is the reliability of the tool-use pipeline, meaning whether the LLM parses the request, calls the solver, and reports the result faithfully without introducing an error. Under the solver-grounded principle the agent inherits the convex solver's feasibility and optimality guarantees, so the quantity worth reporting is how often the LLM breaks that inheritance, which across these 20 days is never. The LLM-only baseline often violates the hard site-peak and other constraints and incurs higher unmet demand. GPT-4o underdelivers $7.5\times$ more unmet energy and Sonnet 4.6 $9.5\times$ more, both exceed the 50\,kW site cap, and their lower cost is an artifact of those violations. {EVAgent} completes within three tool-call rounds per day, about 30\,s (3 to 5 calls, 2 to 5k tokens each) against $<$1\,s for the direct solver. That overhead buys what CVXPY alone cannot: parsing ambiguous natural-language input, multi-turn what-if exploration through the \texttt{disabled\_chargers}, \texttt{site\_cap\_kw}, and \texttt{extra\_sessions} parameters, and faithful reporting where every claim traces to a tool return value. A deterministic parser restricted to structured inputs could not.

\subsection{Power Flow Agent}
\label{sec:pfagent}
\paragraph*{\textbf{Problem Description}}
AC power flow analysis computes the steady-state operating point of an $N$-bus power network given a load-generation injection pattern. The nodal balance equations for active power $P_i$ and reactive power $Q_i$ are
 \begin{align}
 P_i &= V_i \sum_{j=1}^{N} V_j
   \bigl(G_{ij}\cos(\theta_i-\theta_j)
         + B_{ij}\sin(\theta_i-\theta_j)\bigr), \label{eq:pf_p}\\
 Q_i &= V_i \sum_{j=1}^{N} V_j
   \bigl(G_{ij}\sin(\theta_i-\theta_j)
         - B_{ij}\cos(\theta_i-\theta_j)\bigr), \label{eq:pf_q}
 \end{align}
 with operating limits $V_i^{\min} \le V_i \le V_i^{\max}$ on the bus voltage magnitude and $|S_{ij}| \le S_{ij}^{\max}$ on the branch apparent-power flow. Here $V_i$ and $\theta_i$ are the voltage magnitude and angle at bus~$i$; $V_i^{\min}$ and $V_i^{\max}$ are the lower and upper voltage-magnitude limits (e.g., $0.95$ and $1.05$~p.u.); $S_{ij}$ is the apparent-power flow on the branch connecting buses~$i$ and~$j$, and $S_{ij}^{\max}$ is its thermal (MVA) rating. The voltage angles $\theta_i$ are not box-constrained. Only the slack-bus angle is fixed as the reference ($\theta_{\mathrm{ref}}=0$). Angle differences are limited indirectly through the branch-flow limits. Finally, $G_{ij} + j B_{ij}$ is the $(i,j)$ element of the bus admittance matrix $\mathbf{Y}_{\text{bus}}$, with $G_{ij}$ and $B_{ij}$ being the conductance and susceptance, respectively.
 The design question motivating Power Flow Agent (PFAgent) is whether the LLM should solve these equations internally or interpret the request and delegate computation to a solver. PFAgent takes the second route, consistent with the solver-grounded rule of Section~\ref{sec:workflow}: given a natural-language query and current network state $\mathcal{N}$, the agent selects and sequences tool invocations $\{t_1,\ldots,t_R\}$, executes them, and synthesizes the results into a natural-language response. The state $\mathcal{N}$ evolves as the user modifies loads, disconnects branches, or applies remedial actions.
  \paragraph*{\textbf{Strategy}}
  PFAgent has four components: (1) a \textit{Streamlit frontend} \cite{streamlit} managing user interaction, session state, and undo/redo snapshots, (2) an \textit{agentic layer} with a multi-turn conversation engine and tool dispatcher that routes function calls to validated Python implementations, (3) a \textit{solver backend} delegating AC power flow to the PandaPower~\cite{thurner2018pandapower} Newton-Raphson solver and providing N-1 contingency analysis and remedial action generation, and (4) a \textit{visualization layer} rendering topology, voltage heat maps, and flow diagrams (Fig.~S2 in the Supplementary Material). The loop terminates when the LLM returns plain text or hits the round limit (default: 8 rounds, 40 messages). Both LLM-only and PFAgent use the same test cases and metrics. The only difference is that PFAgent can call the PandaPower toolchain, whereas the LLM-only baseline produces the numerical solution directly from the prompt. Fig.~\ref{fig:pf_trajectory} shows a representative PFAgent trajectory on IEEE-14.

\begin{figure}[ht]
\centering
\begin{minipage}{0.94\linewidth}
\begin{promptfigure}{PFAgent Trajectory,  IEEE-14 Power Flow}
\scriptsize
{\ttfamily\scriptsize
\rolehl{User:} Load the IEEE-14 bus system, run AC power flow, and report any voltage violations.\par
\contexthl{Tool calls:} load\_case(case=14) $\to$ run\_powerflow() $\to$ check\_voltage\_violations(v\_min=0.95, v\_max=1.05)\par
\retrievedhl{Tool output:} \{converged: true, iterations: 4, over\_voltage\_buses: [\{bus: 5, v\_pu: 1.070\}, \{bus: 6, v\_pu: 1.062\}, \{bus: 7, v\_pu: 1.090\}], total\_violations: 3, under\_voltage\_buses: []\}\par
\outputhl{Agent:} ``Power flow converged in 4 Newton-Raphson iterations. Three voltage violations detected: Bus~5 (1.070\,p.u.), Bus~6 (1.062\,p.u.), Bus~7 (1.090\,p.u.) - all above the 1.05\,p.u.\ upper limit. No under-voltage violations.''\par
}
\end{promptfigure}
\end{minipage}
\caption{PFAgent on IEEE-14: it loads the case, calls tools, and reports three over-voltage violations. All voltage magnitudes and convergence status come directly from PandaPower; the LLM only synthesizes them.}
\label{fig:pf_trajectory}
\end{figure}

\paragraph*{\textbf{Results}}
We compare two execution modes across three models (Claude Opus 4.7, GPT-5.4, GPT-5.5) on IEEE 14-, 30-, 57-, and 118-bus systems: \textit{LLM-only} (case data as prompt, no solver access) and the proposed \textit{PFAgent}. Because these standard cases likely appear in LLM training data, we randomly perturb loads and generator setpoints around the base MATPOWER case to mitigate memorization and probe generalization, with scalar $k$ controlling the perturbation magnitude. Table~\ref{tab:pf_results} uses $k=1$ with 40 random seeds per test system.

The evaluation uses a three-tier metric set. \textbf{i) State estimation:} voltage-magnitude MAE ($V_{\mathrm{MAE}}$, p.u.), voltage-magnitude RMSE, voltage-angle RMSE (degrees), and maximum voltage error. \textbf{ii) Flow estimation:} branch active-power MAE ($F_{\mathrm{MAE}}$, MW), branch active-power RMSE, 95th-percentile flow error, and branch-loading RMSE (\%). \textbf{iii) Physical consistency:} bus-level active-power balance residual (KCL inconsistency), summarized by $B_{\mathrm{mean}}$. For LLM-only, $B_{\mathrm{mean}} = \frac{1}{N}\sum_i |\Delta P_i|$ from the model-reported flows, with $\Delta P_i = P_i^{\mathrm{inj}} - \sum_{j:(i,j)\in\mathcal{E}} P_{ij}$, where $N$ is the bus count, $P_i^{\mathrm{inj}}$ the net injection at bus~$i$, and $P_{ij}$ the flow on branch $(i,j)\in\mathcal{E}$. For PFAgent it is computed on the full-precision PandaPower solution before JSON rounding. All metrics use ground-truth solutions from the PandaPower Newton-Raphson solver for each perturbed case.

Table~\ref{tab:pf_results} reports one representative metric per tier ($V_{\mathrm{MAE}}$, $F_{\mathrm{MAE}}$, $B_{\mathrm{mean}}$). Because PFAgent delegates computation to PandaPower rather than solving itself, near-solver voltage and flow errors are the expected outcome across both the Claude and GPT families. The quantity worth reading is whether the orchestration preserves the solver numbers, and it does, with bus-balance residuals far below those of the LLM-only baseline. The LLM-only baseline must produce the full numerical solution from the prompt alone, exposing its limited reliability for multi-step computation. $B_{\mathrm{mean}}$ also serves as a faithfulness check: large for the LLM-only baseline because its self-reported flows violate KCL, and below $10^{-6}$ for PFAgent because every reported value traces to the PandaPower solution.

\begin{table}[!ht]
\centering
\caption{PFAgent vs.\ LLM-only on IEEE test systems. $V_{\mathrm{MAE}}$: voltage MAE (p.u.); $F_{\mathrm{MAE}}$: branch active-power MAE (MW); $B_{\mathrm{mean}}$: mean bus-level KCL residual (MW). $k=1$ perturbation, 40 paired seeds.}
\label{tab:pf_results}
\scriptsize
\renewcommand{\arraystretch}{1.15}
\setlength{\tabcolsep}{4pt}
\resizebox{\textwidth}{!}{%
\begin{tabular}{llcccccccccccc}
\toprule
& & \multicolumn{3}{c}{\textbf{IEEE 14-bus}}
  & \multicolumn{3}{c}{\textbf{IEEE 30-bus}}
  & \multicolumn{3}{c}{\textbf{IEEE 57-bus}}
  & \multicolumn{3}{c}{\textbf{IEEE 118-bus}} \\
\cmidrule(lr){3-5}\cmidrule(lr){6-8}\cmidrule(lr){9-11}\cmidrule(lr){12-14}
\textbf{Model} & \textbf{Method}
  & $V_{\mathrm{MAE}}$ & $F_{\mathrm{MAE}}$ & $B_{\mathrm{mean}}$
  & $V_{\mathrm{MAE}}$ & $F_{\mathrm{MAE}}$ & $B_{\mathrm{mean}}$
  & $V_{\mathrm{MAE}}$ & $F_{\mathrm{MAE}}$ & $B_{\mathrm{mean}}$
  & $V_{\mathrm{MAE}}$ & $F_{\mathrm{MAE}}$ & $B_{\mathrm{mean}}^{\dagger}$ \\
\midrule
\multirow{2}{*}{opus-4.7}
  & LLM-only & 0.00682 & 5.23 & 8.58 & 0.00888 & 10.93 & 12.21 & 0.0991 & 31.10 & 38.60 & 0.00748 & 67.83 & 95.19 \\
  & PFAgent  & $\bf{1.65{\times}10^{-5}}$ & \bf{0.00251} & $\bf{<10^{-6}}$ & $\bf{1.92{\times}10^{-5}}$ & \bf{0.00241} & $\bf{<10^{-6}}$ & $\bf{2.20{\times}10^{-5}}$ & \bf{0.00247} & $\bf{<10^{-6}}$ & $\bf{8.55{\times}10^{-6}}$ & \bf{0.00157} & $\bf{<10^{-6}}$ \\
\midrule
\multirow{2}{*}{gpt-5.4}
  & LLM-only & 0.00241 & 2.29 & 1.24 & 0.00738 & 7.29 & 9.11 & 0.1000 & 31.84 & 33.68 & 0.00831 & 53.66 & 60.66 \\
  & PFAgent  & $\bf{1.65{\times}10^{-5}}$ & \bf{0.00251} & $\bf{<10^{-6}}$ & $\bf{1.92{\times}10^{-5}}$ & \bf{0.00241} & $\bf{<10^{-6}}$ & $\bf{3.96{\times}10^{-5}}$ & \bf{0.00247} & $\bf{<10^{-6}}$ & $\bf{1.37{\times}10^{-5}}$ & \bf{0.00250} & $\bf{<10^{-6}}$ \\
\midrule
\multirow{2}{*}{gpt-5.5}
  & LLM-only & 0.00153 & 3.13 & 2.17 & 0.00647 & 6.43 & 8.64 & 0.0874 & 40.64 & 44.38 & 0.00739 & 98.79 & 83.05 \\
  & PFAgent  & $\bf{1.65{\times}10^{-5}}$ & \bf{0.00251} & $\bf{<10^{-6}}$ & $\bf{1.92{\times}10^{-5}}$ & \bf{0.00241} & $\bf{<10^{-6}}$ & $\bf{2.20{\times}10^{-5}}$ & \bf{0.00247} & $\bf{<10^{-6}}$ & $\bf{1.37{\times}10^{-5}}$ & \bf{0.00250} & $\bf{<10^{-6}}$ \\
\bottomrule
\end{tabular}}\\[4pt]
\begin{minipage}{\textwidth}
\scriptsize
Lower values are better for all three metrics. For LLM-only, $B_{\mathrm{mean}}$ is computed from the model-reported branch flows. $^{\dagger}$ For PFAgent, $B_{\mathrm{mean}}$ is computed on the full-precision PandaPower backend solution before JSON rounding; recomputing it from the rounded JSON-formatted answer gives residuals on the order of $10^{-3}$ MW, reflecting serialization precision rather than violation of the underlying AC power-flow equations.
\end{minipage}
\end{table}

\subsection{GridDebugAgent: LLM-Assisted Contingency Diagnosis}
\label{sec:griddebug}

\paragraph*{\textbf{Problem Description}}
Power grid operators face failures such as non-convergent load flow, voltage violations (outside 0.95--1.05~p.u.), and thermal overloads ($>$100\% line loading). Diagnosing root causes and finding corrective actions requires reasoning over network topology, running what-if simulations, and verifying repairs, tasks well suited to LLM agents. We implement \textit{GridDebugAgent}, a solver-grounded ReAct agent~\cite{yao2023react} that iteratively diagnoses failures and executes remediation actions on electrical networks using the PandaPower library~\cite{thurner2018pandapower}.
\paragraph*{\textbf{Strategy}}
GridDebugAgent has three components: (1) a \textit{preprocessing pipeline} with an evidence collector and rule engine that extracts network metrics and classifies failures, (2) a \textit{ReAct agent core} using GPT-4o (temp.\ 0.3, max 50 iterations) with OpenAI function calling, and (3) a tool suite covering five categories: query, simulation, diagnostic, grid actions, and memory management. GPT-4o is the backbone because the agent is built on OpenAI function calling, for which GPT-4o offered strong tool-use support when the system was built. The design is model-agnostic, and the PFAgent and EVAgent case studies exercise the same pattern across other frontier models. The grid-action tools include generation adjustment, load curtailment, line/shunt switching, and voltage control, while memory tools support network snapshots and conversation logging. Given a scenario, GridDebugAgent first applies diagnostic rules, then runs a diagnose$\to$act$\to$verify loop: it queries network state, proposes corrective actions (e.g., adjusting generator setpoints), executes them, re-runs power flow, and checks for remaining violations until the network is secure or max iterations are reached. Figure~\ref{fig:griddebug_trajectory} shows a successful 2-round repair on a line-outage contingency.
\begin{figure}[ht]
\centering
\begin{minipage}{0.94\linewidth}
\begin{promptfigure}{GridDebugAgent Iterative Repair Trajectory: Line Outage on IEEE-14 System}
\scriptsize
{\ttfamily\scriptsize
\rolehl{User:} Disconnect line 2-3 and report any voltage violations or overloads. \par
\reasonhl{Agent (Round 1):} Disconnect line, run power flow, check violations. \par
\contexthl{Tool calls:} save\_network\_snapshot(label=``base'') $\to$ switch\_line(line\_index=2, in\_service=false) $\to$ run\_power\_flow(algorithm=``nr'') $\to$ check\_voltage\_violations(v\_min=0.95, v\_max=1.05) \par
\retrievedhl{Tool output:} \{converged: true, over\_voltage\_buses: [\{bus: 5, v\_pu: 1.070\}, \{bus: 7, v\_pu: 1.090\}], under\_voltage\_buses: [], total\_violations: 2\} \par
\reasonhl{Agent (Round 2):} Over-voltage at buses 5 and 7 --- reduce generator setpoints. 
\contexthl{Tool calls:} adjust\_generation(gen\_index=0, gen\_type=``gen'', vm\_pu\_new=1.05) $\to$ adjust\_generation(gen\_index=1, gen\_type=``gen'', vm\_pu\_new=1.05) $\to$ run\_power\_flow(algorithm=``nr'') $\to$ check\_voltage\_violations(v\_min=0.95, v\_max=1.05) \par
\retrievedhl{Tool output:} \{converged: true, over\_voltage\_buses: [], under\_voltage\_buses: [], total\_violations: 0\} \par
\outputhl{Final output:} ``\goodmark\ Network repaired. Line 2--3 outage caused over-voltage at buses 5 (1.07\,p.u.) and 7 (1.09\,p.u.). Reducing generator setpoints to 1.05\,p.u.\ eliminated all violations. Network converges within limits.''
}
\end{promptfigure}
\end{minipage}
\caption{A GridDebugAgent repair trajectory (reasoning--act--verify loop): it applies a line-outage contingency (Round 1), detects over-voltage at buses 5 and 7, lowers generator setpoints (Round 2), re-runs power flow, and confirms zero violations. All values come from PandaPower.}
\label{fig:griddebug_trajectory}
\end{figure}

The LLM-only baseline tests a single-step structured repair without GridDebugAgent's iterative diagnose--repair--verify loop. It receives the same repair action primitives (\texttt{adjust\_generation}, \texttt{curtail\_load}, \texttt{scale\_all\_loads}, \texttt{switch\_element}, \texttt{add\_shunt\_compensation}, \texttt{adjust\_voltage\_setpoint}, \texttt{no\_action}) but no access to trusted tools, simulation, or memory. Rather than free-text advice, it must call \texttt{propose\_repair\_action}, which returns a JSON action with arguments and rationale. The system validates the action, applies it to the PandaPower network via \texttt{ModificationTools}, and runs power flow once to evaluate the result.
\paragraph*{\textbf{Results}} We evaluate 39 failure scenarios across IEEE 14-, 30-, and 57-bus systems for both the LLM-only baseline and GridDebugAgent. Each network runs 13 preset scenarios (4 non-convergence, 3 voltage, 3 thermal, 2 contingency, 1 normal base case) spanning line outages (N-1), load/generation imbalances, equipment failures, parameter violations, and the unmodified base network. The normal case is included because strict limits may still flag violations. Table~\ref{tab:griddebug_results} reports repair rate (converged, zero violations), improvement rate (converged, strictly fewer violations than initially), feasibility rate, and the initial-to-final violation count.

\begin{table}[!ht]
\centering
\caption{LLM-only vs.\ GridDebugAgent on IEEE networks. \textbf{Rep.}/\textbf{Imp.}/\textbf{Feas.}: runs that converge with zero violations / with fewer violations than initially / at all; \textbf{Viol.}: initial$\rightarrow$final count. Best per metric in \textbf{bold}.}
\label{tab:griddebug_results}
\scriptsize
\setlength{\tabcolsep}{4pt}
\renewcommand{\arraystretch}{1.15}
\begin{tabular}{l *{12}{c}}
\toprule
\multirow{2}{*}{\textbf{Method}}
& \multicolumn{4}{c}{\textbf{IEEE-14}}
& \multicolumn{4}{c}{\textbf{IEEE-30}}
& \multicolumn{4}{c}{\textbf{IEEE-57}} \\
\cmidrule(lr){2-5}\cmidrule(lr){6-9}\cmidrule(lr){10-13}
& \textbf{Rep.} & \textbf{Imp.} & \textbf{Feas.} & \textbf{Viol.}
& \textbf{Rep.} & \textbf{Imp.} & \textbf{Feas.} & \textbf{Viol.}
& \textbf{Rep.} & \textbf{Imp.} & \textbf{Feas.} & \textbf{Viol.} \\
\midrule
LLM-only
& 0.0 & 53.8 & 84.6 & 48$\rightarrow$36
& 23.1 & 46.2 & 76.9 & 128$\rightarrow$122
& 0.0 & 30.8 & 61.5 & 321$\rightarrow$317 \\
GridDebugAgent
& \textbf{61.5} & \textbf{84.6} & \textbf{100.0} & \textbf{48$\rightarrow$8}
& \textbf{46.2} & \textbf{69.2} & \textbf{100.0} & \textbf{128$\rightarrow$67}
& \textbf{23.1} & \textbf{53.8} & \textbf{69.2} & \textbf{321$\rightarrow$162} \\
\bottomrule
\end{tabular}
\end{table}

GridDebugAgent achieves a 43.6\% repair rate (17/39 scenarios) and cuts total violations from 497 to 237, a 52.3\% aggregate reduction, so even partial repairs often yield useful corrective actions. Repair success is measured by re-running PandaPower on the post-action network and counting violations, not from the agent's narrative, so a fabricated ``fixed'' claim does not count and unrepaired scenarios surface their remaining violations rather than hide them. Performance degrades with network size: repair rates of 61.5\%, 46.2\%, and 23.1\% on IEEE-14, -30, and -57 (improvement rates 84.6\%, 69.2\%, 53.8\%), with average tool calls rising from 13.0 to 42.5 as larger networks exceed the iteration budget. The outputs are structurally and semantically consistent with X-GridAgent~\cite{wen2025xgridagent} (e.g., identical bus-index sequences in shortest-path queries, see Table~S3 in the Supplementary Material), confirming alignment on standard analysis tasks even though our contingency-repair benchmark targets a different setting. GridDebugAgent uses the same reasoning-act-verify loop without contingency-specific prompting: the LLM proposes actions from solver-grounded observations, and verification gates the next step.

\subsection{Summary and Indications}
\label{subsec:eval-lessons}
All four case studies instantiate the evaluation protocol proposed in Section~\ref{sec:evaluation}, which are further summarized in Supplementary Material (Table~S1). Each case study reports the task-specific metrics chosen within each of the four groups, plus whether an LLM-only versus solver-grounded comparison is reported. No single case study uses every metric. Each selects the subset that matches its task structure. Wind forecasting is the one case with no external solver: the LLM serves as the prediction engine itself. The solver-grounded correctness layer of Section~\ref{subsec:eval-correctness} then has no tool boundary to check and degrades to \emph{output-level} verification: JSON schema validity and range clipping to the physical $[0,1500]$\,kW envelope. The other three cases instantiate the full constraint-satisfaction checks against a trusted solver. Together, the four groups characterize a solver-grounded system from accuracy through efficiency and surface failure modes that task utility alone would hide.

\section{Conclusion and Discussions}
\label{sec:conclusion}
This tutorial paper presents an implementation-grounded tutorial on LLM-based and agentic systems for smart grids. Written for a power-systems audience, it reviews fundamental building blocks: prompting strategies and agentic architectures, with their memory, tool interfaces, and orchestration frameworks. On this foundation we propose the \emph{solver-grounded design rule}, stating that a numerical result is reported only when it originates from a trusted tool and passes explicit verification. A four-group evaluation protocol spanning task utility, solver-grounded correctness, faithfulness and safe failure, and cost and latency is introduced. We instantiate both the rule and the protocol in four end-to-end case studies: wind power forecasting, EV charging scheduling, power flow analysis, and contingency diagnosis, each comparing an LLM-only baseline against its solver-grounded counterpart on identical data and metrics. A consistent pattern emerged. When the LLM is the interface and a trusted tool is the source of truth, the agent inherits the tool's correctness. When the LLM is asked to compute, it does not. When no solver exists, as in forecasting, verification degrades from physical feasibility to output-level schema and range checks.

Three limitations frame these findings. First, the case studies use off-the-shelf frontier LLMs without domain adaptation, so the comparisons measure prompting and architecture rather than fine-tuning. Second, they cover four representative tasks, not the full design space. Real-time control, market settlement, and protection coordination remain open. Third, the protocol audits each system against its own solver-grounded counterpart rather than comparing two agents head-to-head. The most consequential future work follows from these gaps. Reusable verification libraries could expose per-unit, power-balance, and convergence checks behind a typed interface. Repair policies could turn the safe-failure path into a recoverable one without weakening verification. Verifiable-reward fine-tuning~\cite{lambert2024tulu3} could reuse verification-passed trajectories to train smaller tool-calling agents that run inside utility networks, keeping operational data on premises. None of these replace solvers. They reorganize the work around them by parsing the request, retrieving context, sequencing tool calls, and explaining a verified result. The operational impact of these systems, in our reading, will come not from new numerical capability but from compressing the engineering hours between a question an operator asks in natural language and an answer a control system can act on.

\bibliographystyle{elsarticle-num}
\bibliography{refs}

@inproceedings{yao2023react,
  author    = {Shunyu Yao and Jeffrey Zhao and Dian Yu and Nan Du and Izhak Shafran and Karthik Narasimhan and Yuan Cao},
  title     = {{ReAct}: Synergizing Reasoning and Acting in Language Models},
  booktitle = {International Conference on Learning Representations (ICLR)},
  year      = {2023}
}

@inproceedings{schick2023toolformer,
  author    = {Timo Schick and Jane Dwivedi-Yu and Roberto Dess{\`i} and Roberta Raileanu and Maria Lomeli and Eric Hambro and Luke Zettlemoyer and Nicola Cancedda and Thomas Scialom},
  title     = {{Toolformer}: Language Models Can Teach Themselves to Use Tools},
  booktitle = {Advances in Neural Information Processing Systems (NeurIPS)},
  pages     = {68539--68551},
  year      = {2023}
}

@inproceedings{shinn2023reflexion,
  author    = {Noah Shinn and Federico Cassano and Edward Berman and Ashwin Gopinath and Karthik Narasimhan and Shunyu Yao},
  title     = {{Reflexion}: Language Agents with Verbal Reinforcement Learning},
  booktitle = {Advances in Neural Information Processing Systems (NeurIPS)},
  year      = {2023}
}

@article{wu2023autogen,
  author    = {Qingyun Wu and Gagan Bansal and Jieyu Zhang and Yiran Wu and Beibin Li and Erkang Zhu and Li Jiang and Xiaoyun Zhang and Shaokun Zhang and Jiale Liu and Ahmed Hassan Awadallah and Ryen W. White and Doug Burger and Chi Wang},
  title     = {{AutoGen}: Enabling Next-Gen {LLM} Applications via Multi-Agent Conversation Framework},
  journal   = {arXiv preprint arXiv:2308.08155},
  year      = {2023}
}

@inproceedings{ouyang2022training,
  author    = {Long Ouyang and Jeff Wu and Xu Jiang and Diogo Almeida and Carroll L. Wainwright and Pamela Mishkin and Chong Zhang and Sandhini Agarwal and Katarina Slama and Alex Ray and others},
  title     = {Training Language Models to Follow Instructions with Human Feedback},
  booktitle = {Advances in Neural Information Processing Systems (NeurIPS)},
  pages     = {27730--27744},
  year      = {2022}
}

@inproceedings{brown2020language,
  author    = {Tom B. Brown and Benjamin Mann and Nick Ryder and Melanie Subbiah and Jared Kaplan and Prafulla Dhariwal and Arvind Neelakantan and Pranav Shyam and Girish Sastry and Amanda Askell and others},
  title     = {Language Models are Few-Shot Learners},
  booktitle = {Advances in Neural Information Processing Systems (NeurIPS)},
  pages     = {1877--1901},
  year      = {2020}
}

@article{liu2023pretrain,
  author    = {Pengfei Liu and Weizhe Yuan and Jinlan Fu and Zhengbao Jiang and Hiroaki Hayashi and Graham Neubig},
  title     = {Pre-train, Prompt, and Predict: A Systematic Survey of Prompting Methods in Natural Language Processing},
  journal   = {ACM Computing Surveys},
  volume    = {55},
  number    = {9},
  pages     = {195:1--195:35},
  year      = {2023}
}

@article{schulhoff2024prompt,
  author    = {Sander Schulhoff and Michael Ilie and Nishant Balepur and Konstantine Kahadze and Amanda Liu and Chenglei Si and others},
  title     = {The Prompt Report: A Systematic Survey of Prompt Engineering Techniques},
  journal   = {arXiv preprint arXiv:2406.06608},
  year      = {2024}
}

@inproceedings{wei2022chain,
  author    = {Jason Wei and Xuezhi Wang and Dale Schuurmans and Maarten Bosma and Brian Ichter and Fei Xia and Ed H. Chi and Quoc V. Le and Denny Zhou},
  title     = {Chain-of-Thought Prompting Elicits Reasoning in Large Language Models},
  booktitle = {Advances in Neural Information Processing Systems (NeurIPS)},
  pages     = {24824--24837},
  year      = {2022}
}

@inproceedings{lewis2020rag,
  author    = {Patrick Lewis and Ethan Perez and Aleksandra Piktus and Fabio Petroni and Vladimir Karpukhin and Naman Goyal and Heinrich K{\"u}ttler and Mike Lewis and Wen-tau Yih and Tim Rockt{\"a}schel and Sebastian Riedel and Douwe Kiela},
  title     = {Retrieval-Augmented Generation for Knowledge-Intensive {NLP} Tasks},
  booktitle = {Advances in Neural Information Processing Systems (NeurIPS)},
  pages     = {9459--9474},
  year      = {2020}
}

@article{erdogan2025planact,
  author    = {Lutfi Eren Erdogan and Nicholas Lee and Sehoon Kim and Suhong Moon and Hiroki Furuta and Gopala Anumanchipalli and Kurt Keutzer and Amir Gholami},
  title     = {Plan-and-Act: Improving Planning of Agents for Long-Horizon Tasks},
  journal   = {arXiv preprint arXiv:2503.09572},
  year      = {2025}
}

@inproceedings{ichter2023saycan,
  author    = {Michael Ahn and Anthony Brohan and Noah Brown and Yevgen Chebotar and Omar Cortes and Byron David and others},
  title     = {Do As I Can, Not As I Say: Grounding Language in Robotic Affordances},
  booktitle = {Conference on Robot Learning (CoRL)},
  pages     = {287--318},
  year      = {2023}
}

@inproceedings{liu2023hla,
  author    = {Jijia Liu and Chao Yu and Jiaxuan Gao and Yuqing Xie and Qingmin Liao and Yi Wu and Yu Wang},
  title     = {{LLM}-Powered Hierarchical Language Agent for Real-time Human-AI Coordination},
  booktitle = {Proceedings of the 23rd International Conference on Autonomous Agents and Multiagent Systems (AAMAS)},
  pages     = {1219--1228},
  year      = {2024}
}

@inproceedings{madaan2023selfrefine,
  author    = {Aman Madaan and Niket Tandon and Prakhar Gupta and Skyler Hallinan and Luyu Gao and Sarah Wiegreffe and Uri Alon and Nouha Dziri and Shrimai Prabhumoye and Yiming Yang and others},
  title     = {Self-Refine: Iterative Refinement with Self-Feedback},
  booktitle = {Advances in Neural Information Processing Systems (NeurIPS)},
  year      = {2023}
}

@inproceedings{park2023generative,
  author    = {Joon Sung Park and Joseph C. O'Brien and Carrie J. Cai and Meredith Ringel Morris and Percy Liang and Michael S. Bernstein},
  title     = {Generative Agents: Interactive Simulacra of Human Behavior},
  booktitle = {Proceedings of the 36th Annual ACM Symposium on User Interface Software and Technology (UIST)},
  year      = {2023}
}

@article{packer2023memgpt,
  author    = {Charles Packer and Sarah Wooders and Kevin Lin and Vivian Fang and Shishir G. Patil and Ion Stoica and Joseph E. Gonzalez},
  title     = {{MemGPT}: Towards {LLMs} as Operating Systems},
  journal   = {arXiv preprint arXiv:2310.08560},
  year      = {2023}
}

@article{ji2023hallucination,
  author    = {Ziwei Ji and Nayeon Lee and Rita Frieske and Tiezheng Yu and Dan Su and Yan Xu and Etsuko Ishii and Yejin Bang and Delong Chen and Wenliang Dai and Ho Shu Chan and Andrea Madotto and Pascale Fung},
  title     = {Survey of Hallucination in Natural Language Generation},
  journal   = {ACM Computing Surveys},
  volume    = {55},
  number    = {12},
  pages     = {248:1--248:38},
  year      = {2023}
}

@inproceedings{es2024ragas,
  author    = {Shahul Es and Jithin James and Luis Espinosa-Anke and Steven Schockaert},
  title     = {{RAGAS}: Automated Evaluation of Retrieval Augmented Generation},
  booktitle = {Proceedings of the 18th Conference of the European Chapter of the Association for Computational Linguistics (EACL): System Demonstrations},
  pages     = {150--158},
  year      = {2024}
}

@inproceedings{geifman2017selective,
  author    = {Yonatan Geifman and Ran El-Yaniv},
  title     = {Selective Classification for Deep Neural Networks},
  booktitle = {Advances in Neural Information Processing Systems (NeurIPS)},
  pages     = {4878--4887},
  year      = {2017}
}

@article{yao2024taubench,
  author    = {Shunyu Yao and Noah Shinn and Pedram Razavi and Karthik Narasimhan},
  title     = {{$\tau$-bench}: A Benchmark for Tool-Agent-User Interaction in Real-World Domains},
  journal   = {arXiv preprint arXiv:2406.12045},
  year      = {2024}
}

@inproceedings{wei2022finetuned,
  author    = {Jason Wei and Maarten Bosma and Vincent Y. Zhao and Kelvin Guu and Adams Wei Yu and Brian Lester and Nan Du and Andrew M. Dai and Quoc V. Le},
  title     = {Finetuned Language Models Are Zero-Shot Learners},
  booktitle = {International Conference on Learning Representations (ICLR)},
  year      = {2022}
}

@article{chen2021codex,
  author    = {Mark Chen and Jerry Tworek and Heewoo Jun and Qiming Yuan and Henrique Ponde de Oliveira Pinto and Jared Kaplan and others},
  title     = {Evaluating Large Language Models Trained on Code},
  journal   = {arXiv preprint arXiv:2107.03374},
  year      = {2021}
}

@article{boiko2023autonomous,
  author    = {Daniil A. Boiko and Robert MacKnight and Ben Kline and Gabe Gomes},
  title     = {Autonomous Chemical Research with Large Language Models},
  journal   = {Nature},
  volume    = {624},
  number    = {7992},
  pages     = {570--578},
  year      = {2023}
}

@article{singhal2023medpalm,
  author    = {Karan Singhal and Shekoofeh Azizi and Tao Tu and S. Sara Mahdavi and Jason Wei and Hyung Won Chung and others},
  title     = {Large Language Models Encode Clinical Knowledge},
  journal   = {Nature},
  volume    = {620},
  number    = {7972},
  pages     = {172--180},
  year      = {2023}
}

@inproceedings{wu2024timeseries_llm,
  author    = {Nate Gruver and Marc Finzi and Shikai Qiu and Andrew Gordon Wilson},
  title     = {Large Language Models Are Zero-Shot Time Series Forecasters},
  booktitle = {Advances in Neural Information Processing Systems (NeurIPS)},
  year      = {2023}
}

@misc{openai_gpt55_2026,
  author    = {{OpenAI}},
  title     = {{GPT-5.5} Model},
  howpublished= {OpenAI API documentation},
  year      = {2026},
  url       = {https://developers.openai.com/api/docs/models/gpt-5.5}
}

@misc{anthropic_opus48,
  author    = {{Anthropic}},
  title     = {Introducing {Claude Opus 4.8}},
  howpublished= {Anthropic},
  year      = {2026},
  url       = {https://www.anthropic.com/news/claude-opus-4-8}
}

@misc{google_gemini31pro_2026,
  author    = {{Google}},
  title     = {{Gemini 3.1 Pro} (Preview)},
  howpublished= {Google AI for Developers},
  year      = {2026},
  url       = {https://ai.google.dev/gemini-api/docs/models/gemini-3.1-pro-preview}
}

@article{lambert2024tulu3,
  author    = {Nathan Lambert and Jacob Morrison and Valentina Pyatkin and Shengyi Huang and Hamish Ivison and Faeze Brahman and others},
  title     = {{T\"ulu 3}: Pushing Frontiers in Open Language Model Post-Training},
  journal   = {arXiv preprint arXiv:2411.15124},
  year      = {2024}
}

@misc{meta_llama4scout_2026,
  author    = {{Meta AI}},
  title     = {{Llama 4} (Scout)},
  howpublished= {Meta},
  year      = {2025},
  url       = {https://www.llama.com/models/llama-4/}
}

@misc{deepseek_r1_2025,
  author    = {{DeepSeek-AI}},
  title     = {{DeepSeek-R1}: Incentivizing Reasoning Capability in {LLMs} via Reinforcement Learning},
  howpublished= {GitHub repository},
  year      = {2025},
  url       = {https://github.com/deepseek-ai/DeepSeek-R1}
}

@misc{streamlit,
    title = {A Faster Way to Build and Share Data apps},
    url = {https://streamlit.io/}
}

@misc{ERA5,
  author    = {European Centre for Medium-Range Weather Forecasts},
  title     = {ERA5 hourly data on single levels from 1940 to present},
  url       = {https://cds.climate.copernicus.eu/datasets/reanalysis-era5-single-levels?tab=overview}}

@article{zhou2024sdwpf,
  author    = {Jingbo Zhou and Xinjiang Lu and Yixiong Xiao and Jian Tang and Jiantao Su and Yu Li and Ji Liu and Junfu Lyu and Yanjun Ma and Dejing Dou},
  title     = {SDWPF: A dataset for spatial dynamic wind power forecasting over a large turbine array},
  journal   = {Scientific Data},
  volume    = {11},
  number    = {1},
  pages     = {649},
  year      = {2024}
}

@misc{qwen_qwen3instruct2507_2025,
  author    = {{Qwen Team}},
  title     = {{Qwen3-235B-A22B-Instruct-2507}},
  howpublished= {Hugging Face model card},
  year      = {2025},
  url       = {https://huggingface.co/Qwen/Qwen3-235B-A22B-Instruct-2507}
}

@misc{mistral_large3_2025,
  author    = {{Mistral AI}},
  title     = {{Mistral Large 3}},
  howpublished= {Mistral AI},
  year      = {2025},
  url       = {https://mistral.ai/news/mistral-3/}
}

@book{kirschen2024power,
  author    = {Daniel S. Kirschen},
  title     = {Power Systems: Fundamental Concepts and the Transition to Sustainability},
  publisher = {Wiley},
  year      = {2024}
}

@article{thurner2018pandapower,
  author    = {Leon Thurner and Alexander Scheidler and Florian Sch{\"a}fer and Jan-Hendrik Menke and Julian Dollichon and Friederike Meier and Steffen Meinecke and Martin Braun},
  title     = {pandapower---An Open-Source Python Tool for Convenient Modeling, Analysis, and Optimization of Electric Power Systems},
  journal   = {IEEE Transactions on Power Systems},
  volume    = {33},
  number    = {6},
  pages     = {6510--6521},
  year      = {2018}
}

@inproceedings{lee2019acndata,
  author    = {Zachary J. Lee and Tongxin Li and Steven H. Low},
  title     = {{ACN-Data}: Analysis and Applications of an Open {EV} Charging Dataset},
  booktitle = {Proceedings of the Tenth ACM International Conference on Future Energy Systems (e-Energy)},
  pages     = {139--149},
  year      = {2019}
}

@article{diamond2016cvxpy,
  author    = {Steven Diamond and Stephen Boyd},
  title     = {{CVXPY}: A Python-Embedded Modeling Language for Convex Optimization},
  journal   = {Journal of Machine Learning Research},
  volume    = {17},
  number    = {83},
  pages     = {1--5},
  year      = {2016}
}

@article{tang2025timeseries,
  author    = {Hua Tang and Chong Zhang and Mingyu Jin and Qinkai Yu and Zhenting Wang and Xiaobo Jin and Yongfeng Zhang and Mengnan Du},
  title     = {Time Series Forecasting with {LLMs}: Understanding and Enhancing Model Capabilities},
  journal   = {ACM SIGKDD Explorations Newsletter},
  volume    = {26},
  number    = {2},
  pages     = {109--118},
  year      = {2025}
}

@article{takrouri2025distillation,
  author    = {Mohannad Takrouri and Nicol{\'a}s M. Cuadrado and Martin Tak{\'a}{\v c}},
  title     = {Knowledge Distillation from Large Language Models for Household Energy Modeling},
  journal   = {arXiv preprint arXiv:2502.03034},
  year      = {2025}
}

@inproceedings{huang2024foundation,
  author    = {Chenghao Huang and Siyang Li and Ruohong Liu and Hao Wang and Yize Chen},
  title     = {Large Foundation Models for Power Systems},
  booktitle = {2024 IEEE Power \& Energy Society General Meeting (PESGM)},
  pages     = {1--5},
  year      = {2024}
}

@article{majumder2024electric,
  author    = {Subir Majumder and Lin Dong and Fatemeh Doudi and Yuting Cai and Chao Tian and Dileep Kalathil and Kevin Ding and Anupam A. Thatte and Na Li and Le Xie},
  title     = {Exploring the Capabilities and Limitations of Large Language Models in the Electric Energy Sector},
  journal   = {Joule},
  volume    = {8},
  number    = {6},
  pages     = {1544--1549},
  year      = {2024}
}

@article{huang2024llm_optimization_survey,
  author    = {Sen Huang and Kaixiang Yang and Sheng Qi and Rui Wang},
  title     = {When Large Language Model Meets Optimization},
  journal   = {Swarm and Evolutionary Computation},
  volume    = {90},
  pages     = {101663},
  year      = {2024}
}

@article{hamann2024grid_fms,
  author    = {Hendrik F. Hamann and Blazhe Gjorgiev and Thomas Brunschwiler and Leonardo S. A. Martins and Alban Puech and Anna Varbella and Jonas Weiss and Juan Bernabe-Moreno and others},
  title     = {Foundation Models for the Electric Power Grid},
  journal   = {Joule},
  volume    = {8},
  number    = {12},
  pages     = {3245--3258},
  year      = {2024}
}

@article{liu2025repower,
  author    = {Yu-Xiao Liu and Mengshuo Jia and Yong-Xin Zhang and Jianxiao Wang and Guannan He and Shao-Long Zhong and Zhi-Min Dang},
  title     = {{RePower}: An {LLM}-Driven Autonomous Platform for Power System Data-Guided Research},
  journal   = {Patterns},
  volume    = {6},
  number    = {4},
  pages     = {101211},
  year      = {2025}
}

@article{jin2025gridmind,
  author    = {Hongwei Jin and Kibaek Kim and Jonghwan Kwon},
  title     = {{GridMind}: {LLMs}-Powered Agents for Power System Analysis and Operations},
  journal   = {arXiv preprint arXiv:2509.02494},
  year      = {2025}
}

@article{cheng2025gaia,
  author    = {Yuheng Cheng and Huan Zhao and Xiyuan Zhou and Junhua Zhao and Yuji Cao and Chao Yang and Xinlei Cai},
  title     = {A Large Language Model for Advanced Power Dispatch},
  journal   = {Scientific Reports},
  volume    = {15},
  number    = {1},
  pages     = {8925},
  year      = {2025}
}

@article{cheng2026rag,
  author    = {Yuheng Cheng and Huan Zhao and Dejun Xiang and Zhengwen Zhang and Guolong Liu and Yanli Liu and Junhua Zhao and Xinlei Cai},
  title     = {Power System Operational Reliability Evaluation with Retrieval-Augmented Generation Enhanced Large Language Model},
  journal   = {Energy and AI},
  volume    = {24},
  pages     = {100688},
  year      = {2026}
}

@article{mohammadi2025ed,
  author    = {Sina Mohammadi and Ali Hassan and Rouzbeh Haghighi and Van-Hai Bui and Wencong Su},
  title     = {Large Language Models for Solving Economic Dispatch Problem},
  journal   = {arXiv preprint arXiv:2505.21931},
  year      = {2025}
}

@article{wen2025xgridagent,
  author    = {Yihan Wen and Xin Chen},
  title     = {{X-GridAgent}: An {LLM}-Powered Agentic {AI} System for Assisting Power Grid Analysis},
  journal   = {arXiv preprint arXiv:2512.20789},
  year      = {2025}
}

@article{zhang2025gridagent,
  author    = {Yan Zhang and Ahmad Mohammad Saber and Amr Youssef and Deepa Kundur},
  title     = {{Grid-Agent}: An {LLM}-Powered Multi-Agent System for Power Grid Control},
  journal   = {arXiv preprint arXiv:2508.05702},
  year      = {2025}
}

@article{jia2025feedback,
  author    = {Mengshuo Jia and Zeyu Cui and Gabriela Hug},
  title     = {Enhancing {LLMs} for Power System Simulations: A Feedback-Driven Multi-Agent Framework},
  journal   = {IEEE Transactions on Smart Grid},
  volume    = {16},
  number    = {6},
  pages     = {5556--5572},
  year      = {2025}
}

@article{deng2025gridmodel,
  author    = {K. Deng and Y. Zhou and H. Zeng and Z. Wang and Q. Guo},
  title     = {Power Grid Model Generation Based on the Tool-Augmented Large Language Model},
  journal   = {IEEE Transactions on Power Systems},
  volume    = {40},
  number    = {6},
  pages     = {5487--5490},
  year      = {2025}
}

@article{bernier2025powergraphllm,
  author    = {Fabien Bernier and Jun Cao and Maxime Cordy and Salah Ghamizi},
  title     = {{PowerGraph-LLM}: Novel Power Grid Graph Embedding and Optimization With Large Language Models},
  journal   = {IEEE Transactions on Power Systems},
  volume    = {40},
  number    = {6},
  pages     = {5483--5486},
  year      = {2025}
}

@article{hu2025solverready_uc,
  author    = {Yunkai Hu and Tianqiao Zhao and Meng Yue},
  title     = {From Natural Language to Solver-Ready Power System Optimization: An {LLM}-Assisted, Validation-in-the-Loop Framework},
  journal   = {arXiv preprint arXiv:2508.08147},
  year      = {2025}
}

@article{amjad2025review,
  author    = {Furqan Amjad and Tarmo Kor{\~o}tko and Argo Rosin},
  title     = {Review of {LLMs} Applications in Electrical Power and Energy Systems},
  journal   = {IEEE Access},
  volume    = {13},
  pages     = {150951--150969},
  year      = {2025}
}

@article{madani2025smartgrids,
  author    = {Seyyedreza Madani and Ahmadreza Tavasoli and Zahra Khoshtarash Astaneh and Pierre-Olivier Pineau},
  title     = {Large Language Models Integration in Smart Grids},
  journal   = {Energy Reports},
  volume    = {14},
  pages     = {1562--1577},
  year      = {2025}
}

@article{sarwar2025survey,
  author    = {Muhammad Sarwar and Muhammad Rizwan and Mubushra Aziz and Abdul Rehman Sudais},
  title     = {Large Language Models for Power System Applications: A Comprehensive Literature Survey},
  journal   = {arXiv preprint arXiv:2512.13004},
  year      = {2025}
}

@article{agentic_power_2025,
  author    = {Soham Ghosh and Gaurav Mittal},
  title     = {Agentic {AI} Systems in Electrical Power Systems Engineering: Current State-of-the-Art and Challenges},
  journal   = {arXiv preprint arXiv:2511.14478},
  year      = {2025}
}

@article{zhang2025poweragent,
  author    = {Qian Zhang and Le Xie},
  title     = {{PowerAgent}: A Road Map Toward Agentic Intelligence in Power Systems---Foundation Model, Model Context Protocol, and Workflow},
  journal   = {IEEE Power and Energy Magazine},
  volume    = {23},
  number    = {5},
  pages     = {93--101},
  year      = {2025}
}

@article{xie2026foundation,
  author    = {Le Xie and Qian Zhang and Minlan Yu and Paul L. Joskow and Chanan Singh},
  title     = {Crucial Role of Foundation Models in Enhancing the Interaction of {AI} and Power Systems},
  journal   = {IEEE Energy Sustainability Magazine},
  volume    = {2},
  number    = {1},
  pages     = {22--28},
  year      = {2026}
}

@article{ibrahim2025cyber_survey,
  author    = {Nourhan Ibrahim and Rasha Kashef},
  title     = {Exploring the Emerging Role of Large Language Models in Smart Grid Cybersecurity: A Survey of Attacks, Detection Mechanisms, and Mitigation Strategies},
  journal   = {Frontiers in Energy Research},
  volume    = {13},
  pages     = {1531655},
  year      = {2025}
}

@inproceedings{mohammadi2025agenteval,
  author    = {Mahmoud Mohammadi and Yipeng Li and Jane Lo and Wendy Yip},
  title     = {Evaluation and Benchmarking of {LLM} Agents: A Survey},
  booktitle = {Proceedings of the 31st ACM SIGKDD Conference on Knowledge Discovery and Data Mining (KDD), V.2},
  pages     = {6129--6139},
  year      = {2025}
}

@article{lin2025agenthallucinations,
  author    = {Xixun Lin and Yucheng Ning and Jingwen Zhang and Yan Dong and Yilong Liu and Yongxuan Wu and others},
  title     = {{LLM}-Based Agents Suffer from Hallucinations: A Survey of Taxonomy, Methods, and Directions},
  journal   = {arXiv preprint arXiv:2509.18970},
  year      = {2025}
}

@article{liu2025dispatch_redesign,
  author    = {Guangyi Liu and Yachen Tang and Wenxin Guo and Jian Dang},
  title     = {Redesigning Power Grid Dispatching Automation Systems by Integrating {LLMs}, {KGs}, and {AI} Agents},
  journal   = {CSEE Journal of Power and Energy Systems},
  volume    = {11},
  number    = {6},
  pages     = {2610--2622},
  year      = {2025}
}

@article{jena2025incontext_mpc,
  author    = {Amit Jena and Fei Ding and Jiyu Wang and Yiyun Yao and Le Xie},
  title     = {{LLM}-Based Adaptive Distribution Voltage Regulation Under Frequent Topology Changes: An In-Context {MPC} Framework},
  journal   = {IEEE Transactions on Smart Grid},
  volume    = {16},
  number    = {5},
  pages     = {4297--4300},
  year      = {2025}
}

@article{ji2025lemad,
  author    = {Xin Ji and Le Zhang and Wenya Zhang and Fang Peng and Yifan Mao and Xingchuang Liao and Kui Zhang},
  title     = {{LEMAD}: {LLM}-Empowered Multi-Agent System for Anomaly Detection in Power Grid Services},
  journal   = {Electronics},
  volume    = {14},
  number    = {15},
  pages     = {3008},
  year      = {2025}
}

@inproceedings{saha2025dragent,
  author    = {Barun Kumar Saha and Aarthi V. and O. D. Naidu},
  title     = {{DrAgent}: An Agentic Approach to Fault Analysis in Power Grids Using Large Language Models},
  booktitle = {2025 International Conference on Artificial Intelligence in Information and Communication (ICAIIC)},
  pages     = {938--945},
  year      = {2025}
}

@article{liao2026zerofewshot,
  author    = {Wenlong Liao and Zhe Yang and Mengshuo Jia and Christian Rehtanz and Jiannong Fang and Fernando Port{\'e}-Agel},
  title     = {Zero-Shot Load Forecasting with Large Language Models},
  journal   = {arXiv preprint arXiv:2411.11350},
  year      = {2024}
}

@article{liu2025cpllmwpf,
  author    = {Yuqi Liu and Weimin Yuan and Weilong Chen and Wenming Li and Han Yang and Yanru Zhang},
  title     = {{CPLLM-WPF}: A Multi-Scale Prompting Framework for Generalizable Wind Power Forecasting with {LLMs}},
  journal   = {Applied Energy},
  volume    = {402},
  pages     = {126912},
  year      = {2025}
}

@article{makroum2025hems,
  author    = {El Makroum, Reda and Sebastian Zwickl-Bernhard and Lukas Kranzl},
  title     = {Agentic {AI} Home Energy Management System: A Large Language Model Framework for Residential Load Scheduling},
  journal   = {Results in Engineering},
  year      = {2026}
}

@article{xiong2026evbehavior,
  author    = {Yi Xiong and Jiamin Ge and Liang Che},
  title     = {{LLM} and Bayesian Network Integrated Simulation Framework for Electric Vehicle User Charging Behaviors},
  journal   = {IEEE Transactions on Smart Grid},
  volume    = {17},
  number    = {3},
  pages     = {2618--2621},
  year      = {2026}
}

@article{zhang2025drev,
  author    = {Zhaoyi Zhang and Chenggang Cui and Ning Yang and Chuanlin Zhang},
  title     = {Research on a Two-Layer Demand Response Framework for Electric Vehicle Users and Aggregators Based on {LLMs}},
  journal   = {arXiv preprint arXiv:2505.20877},
  year      = {2025}
}

@article{zheng2025evplanning,
  author    = {Xinda Zheng and Canchen Jiang and Hao Wang},
  title     = {Large Language Model-Assisted Planning of Electric Vehicle Charging Infrastructure with Real-World Case Study},
  journal   = {Sustainable Energy Technologies and Assessments},
  volume    = {84},
  pages     = {104723},
  year      = {2025}
}

@article{lu2026evoccupancy,
  author    = {Yuhong Lu and Libao Shi},
  title     = {An Enhanced Explanatory Large Language Model-Based Framework for Electric Vehicle Charging Station Occupancy Prediction},
  journal   = {Energy and AI},
  volume    = {24},
  pages     = {100733},
  year      = {2026}
}

@misc{openai2025agentssdk,
  author    = {{OpenAI}},
  title     = {{OpenAI Agents SDK}},
  howpublished= {\url{https://openai.github.io/openai-agents-python/}},
  year      = {2025},
  note      = {Software documentation, accessed May 2026}
}

@misc{microsoft2025semantickernel,
  author    = {{Microsoft}},
  title     = {{Semantic Kernel}},
  howpublished= {\url{https://learn.microsoft.com/en-us/semantic-kernel/}},
  year      = {2025},
  note      = {Software documentation, accessed May 2026}
}

@misc{aws2025strands,
  author    = {{AWS}},
  title     = {{Strands Agents}},
  howpublished= {\url{https://strandsagents.com/}},
  year      = {2025},
  note      = {Software documentation, accessed May 2026}
}

@misc{crewai2025,
  author    = {{CrewAI}},
  title     = {{CrewAI}},
  howpublished= {\url{https://docs.crewai.com/}},
  year      = {2025},
  note      = {Software documentation, accessed May 2026}
}

@misc{langgraph2025,
  author    = {{LangChain}},
  title     = {{LangGraph}},
  howpublished= {\url{https://docs.langchain.com/oss/python/langgraph/overview}},
  year      = {2025},
  note      = {Software documentation, accessed May 2026}
}

@misc{microsoft2025agentframework,
  author    = {{Microsoft}},
  title     = {{Microsoft Agent Framework}},
  howpublished= {\url{https://learn.microsoft.com/en-us/agent-framework/}},
  year      = {2025},
  note      = {Software documentation, accessed May 2026}
}

@misc{anthropic2024mcp,
  author    = {{Anthropic}},
  title     = {{Model Context Protocol (MCP)}},
  howpublished= {\url{https://modelcontextprotocol.io/}},
  year      = {2024},
  note      = {Specification, accessed May 2026}
}

@misc{google2025a2a,
  author    = {{Google}},
  title     = {{Agent2Agent (A2A) Protocol}},
  howpublished= {\url{https://a2a-protocol.org/latest/}},
  year      = {2025},
  note      = {Specification, accessed May 2026}
}

@misc{pichai2026io,
  author       = {{Google}},
  title        = {Google {I/O} 2026: {Sundar} {Pichai}'s Opening Keynote},
  howpublished = {\url{https://blog.google/innovation-and-ai/sundar-pichai-io-2026/}},
  year         = {2026},
  note         = {Reports over 3.2 quadrillion monthly tokens processed across Google surfaces, a 7$\times$ year-over-year increase. Accessed: 2026-07-04}
}

@misc{gartner2025agentic,
  author       = {{Gartner}},
  title        = {Gartner Predicts 40\% of Enterprise Apps Will Feature Task-Specific {AI} Agents by 2026, Up from Less Than 5\% in 2025},
  howpublished = {\url{https://www.gartner.com/en/newsroom/press-releases/2025-08-26-gartner-predicts-40-percent-of-enterprise-apps-will-feature-task-specific-ai-agents-by-2026-up-from-less-than-5-percent-in-2025}},
  year         = {2025},
  note         = {Press release, Stamford, CT, Aug.\ 26, 2025. Accessed: 2026-07-04}
}

\end{document}


\begin{frontmatter}
\title{Supplementary Material\\ LLMs and Agentic AI Systems for Smart Grids: A Tutorial on Architectures and Applications}
\end{frontmatter}

This document collects supplementary material for the main article.

\section{Evaluation Framework Across Case Studies}
Table~S1 maps the four-group evaluation framework onto the four case studies.

\begin{table}[htbp]
\caption{Mapping of the four-group evaluation framework onto the four case studies. Each cell lists the task-specific metric(s) per group. The last column marks whether an LLM-only baseline was run on identical data and metrics.}
\label{stab:eval_mapping}
\centering
\footnotesize
\renewcommand{\arraystretch}{1.2}
\setlength{\tabcolsep}{4pt}
\begin{tabular}{@{}p{0.12\linewidth}p{0.15\linewidth}p{0.26\linewidth}p{0.21\linewidth}p{0.15\linewidth}@{}}
\toprule
\textbf{Case study} & \textbf{Task utility} & \textbf{Solver-grounded} & \textbf{Faithful / safe} & \textbf{Cost / latency} \\
\midrule
Wind forecasting & MAE, RMSE over 3\,h / 6\,h / 48\,h horizons & \emph{No solver}; output-level checks: schema validity, clipping to $[0,1500]$\,kW & JSON schema check with retry & Prompt tokens per horizon \\
EV scheduling & Cost (\$), served~\% & Checker flags (hard + energy shortfall) & All reported numbers traced to tool JSON & Tool calls per day, latency \\
Power-flow agent & VMAE, FMAE, branch-loading RMSE & Newton--Raphson convergence; KCL self-consistency ($B_{\mathrm{mean}}$) & Schema-valid success rate & Rounds under 8-round cap \\
GridDebug & Repair rate, violation reduction & pandapower convergence; feasibility~\% & Numbers sourced only from pandapower output & Iterations, tool calls, runtime \\
\bottomrule
\end{tabular}
\end{table}

\section{Failure Analysis Across Case Studies}
\label{sec:appendix-failures}
Table~\ref{stab:failure_modes} documents the scenarios where each system reaches its limit. EVAgent and PFAgent preserve the solver result on every evaluated case, so their hard failures are confined to the LLM-only baselines. GridDebugAgent is where the agentic framework itself leaves scenarios unsolved, and wind forecasting has no solver, so its failures are prompting-level.

\begin{table}[htbp]
\caption{Observed failure modes across the four case studies, with the specific scenarios and their boundary. }
\label{stab:failure_modes}
\centering
\footnotesize
\renewcommand{\arraystretch}{1.2}
\setlength{\tabcolsep}{4pt}
\begin{tabularx}{\textwidth}{@{}p{0.16\linewidth}p{0.20\linewidth}X@{}}
\toprule
\textbf{Case study} & \textbf{Failure mode} & \textbf{Specifics and boundary} \\
\midrule
Wind forecasting & Prompting fragility & No model--strategy pair dominates across horizons. Claude Sonnet~4.6 with APBF collapses at the 3\,h horizon (Overall 1104.98 versus 366.39 for the naive prompt), and Gemini~3 with the advanced prompt is worse than naive at 3\,h (524.81 versus 147.21). No trusted solver exists, so advanced prompting shapes output without guaranteeing gains. \\
EV scheduling & Residual unmet energy under congestion & The CVXPY optimum itself leaves 27.82\,kWh unmet and 15.80 shortfall flags on high-congestion days, an infeasibility under the 50\,kW site cap rather than an agent error. Hard constraint violations occur only for LLM-only (209--265\,kWh unmet, site-cap breaches). \\
Power flow & Round cap and LLM-only unreliability & PFAgent preserves the solver numbers ($B_{\mathrm{mean}}<10^{-6}$) within the 8-round, 40-message cap. LLM-only is unreliable on larger systems, with VMAE near 0.1\,p.u.\ and KCL residuals up to 95\,MW on IEEE 57--118. \\
Contingency diagnosis & Incomplete repair at scale & GridDebugAgent fully repairs 17 of 39 scenarios. Repair rate falls from 61.5\% to 46.2\% to 23.1\% on IEEE 14/30/57, and 162 of 321 violations remain on IEEE-57 as average tool calls rise from 13.0 to 42.5 and larger networks exceed the iteration budget. \\
\bottomrule
\end{tabularx}
\end{table}

Every failure above is measured, not narrated. The shared checker recomputes each metric on the system the agent produced, so partial repairs and residual violations surface in the reported numbers rather than being masked. This is the safe-failure behavior of Section~\ref{sec:solver-grounded} in practice: the system reports the shortfall instead of fabricating a fix.

\section{Reproducibility}
\label{sec:reproducibility}

All code, data pointers, and experimental settings are released as a self-contained repository with one sub-folder per case study (\texttt{wind-forecasting}, \texttt{ev-scheduling}, \texttt{power-flow-agent}, and \texttt{griddebug-agent}).\footnote{Code and data: \url{https://github.com/d2rojas/LLMs-Agents-For-SmartGrids-Code}} Each sub-folder provides a \texttt{README}, a pinned \texttt{requirements.txt}, the exact models, decoding temperatures, and random seeds, the LLM-only baseline prompts, and the evaluation harness that regenerates the corresponding result table in Section~\ref{sec:casestudies}. Experiments run on Python~3.11 with public datasets (SDWPF~\cite{zhou2024sdwpf}, Caltech ACN-Data~\cite{lee2019acndata}, MATPOWER IEEE cases) and the open-source solvers CVXPY~\cite{diamond2016cvxpy} and PandaPower~\cite{thurner2018pandapower}. The frontier LLMs are accessed through providers' APIs and used off the shelf, without fine-tuning. Each sub-folder installs from its pinned requirements on a clean environment and ships a test suite that exercises the solver and verification paths without API access, so the trusted-tool components can be validated independently of the LLM calls.

\subsection{Case-study setups}
\label{sec:appendix-cases}
\begin{itemize}
\item \textbf{Wind power forecasting} (Section~\ref{sec:casestudies}-A; Table~\ref{tab:wind-complete_results}). Given a 14-day history of SCADA features and an optional 48-hour ERA5 100\,m wind forecast~\cite{ERA5} from Open-Meteo, predict the next 48 hours of active power as $288$ values. The LLM is the predictor, so there is no solver and verification is at the output level (JSON schema validity, clipping to $[0,1500]$\,kW). We compare Naive, Advanced, and APBF prompting against a GRU baseline on the SDWPF dataset~\cite{zhou2024sdwpf} (134 turbines, 10-min, 245 days) at 3/6/48-hour horizons, using Gemini~3 Flash, GPT-5.4, and Claude Sonnet~4.6.
\item \textbf{EV charging scheduling} (Section~\ref{sec:casestudies}-B; Table~\ref{tab:ev_results}). Allocate power to concurrent sessions over a 96-step day under per-session and 50\,kW site caps, minimizing time-of-use cost; the trusted tool is a CVXPY linear program~\cite{diamond2016cvxpy}. We compare the direct solver, LLM-only, and EVAgent (GPT-4o and Claude Sonnet~4.6, temperature~0.7, 5 runs) over 20 days of Caltech ACN-Data, JPL site~\cite{lee2019acndata}, scored by a shared checker for hard violations and energy-shortfall flags.
\item \textbf{Power flow analysis} (Section~\ref{sec:casestudies}-C; Table~\ref{tab:pf_results}). Compute the AC operating point with PandaPower's Newton--Raphson solver~\cite{thurner2018pandapower}. Loads and generator setpoints on the MATPOWER IEEE 14/30/57/118 cases are perturbed ($k{=}1$, $N{=}40$ seeds per system) to probe generalization; LLM-only and PFAgent (Claude Opus~4.7, GPT-5.4, GPT-5.5) are scored on a three-tier metric set (voltage error, branch-flow error, and bus-balance/KCL residual).
\item \textbf{Contingency diagnosis} (Section~\ref{sec:casestudies}-D; Table~\ref{tab:griddebug_results}). Diagnose and repair failures (non-convergence, voltage violations outside $0.95$--$1.05$\,p.u., thermal overloads $>$100\%) through a ReAct diagnose--act--verify loop over PandaPower. GridDebugAgent (GPT-4o, temperature~0.3, $\le$50 iterations) and an LLM-only single-step baseline are run on 39 preset scenarios across IEEE 14/30/57, reporting repair, improvement, feasibility, and violation reduction.
\end{itemize}

\section{Comparison of LLM- and Agentic Systems for Smart Grids}
\label{sec:appendix-comparison}
Table~\ref{tab:systems_comparison} contrasts representative LLM- and agent-based systems along the dimensions this paper emphasizes: the trusted tool each wraps and whether it applies an \emph{explicit} verification gate with safe failure rather than merely relaying solver output. Most prior systems rely on solver results but do not make verification a distinct, reportable step, the gap the solver-grounded rule of Section~\ref{sec:solver-grounded} addresses.

\begin{table}[t]
\centering
\caption{Representative LLM- and agent-based systems for smart grids, compared along the dimensions emphasized in this paper. ``Explicit verification / safe fail'' denotes a dedicated gate that validates the final result and triggers safe failure or repair, beyond relaying solver output: \emph{implicit} means correctness rests on the solver call without a separate gate, \emph{partial} means a validation or feedback step is present but without a safe-failure contract, and \emph{---} marks roadmap or no-solver designs.}
\label{tab:systems_comparison}
\scriptsize
\renewcommand{\arraystretch}{1.2}
\setlength{\tabcolsep}{3pt}
\begin{tabularx}{\textwidth}{@{}l Y l l c@{}}
\toprule
\textbf{System} & \textbf{Task / scope} & \textbf{Architecture} & \textbf{Trusted tool} & \textbf{Explicit verif.\,/\,safe fail} \\
\midrule
GridMind~\cite{jin2025gridmind} & AC OPF, N-1 contingency & Multi-agent & PandaPower & implicit \\
Grid-Agent~\cite{zhang2025gridagent} & Violation detection \& remediation & Planner $+$ validation agent & PandaPower (sandboxed) & partial \\
X-GridAgent~\cite{wen2025xgridagent} & Grid analysis, Q\&A, power flow & Hierarchical (MCP) & PandaPower $+$ KB & implicit \\
Jia et al.~\cite{jia2025feedback} & Power-system analysis & Feedback-driven multi-agent & Daline & partial \\
PowerAgent~\cite{zhang2025poweragent} & Grid operations (roadmap) & Foundation model $+$ MCP & PowerWorld, PandaPower & --- \\
PowerGraph-LLM~\cite{bernier2025powergraphllm} & OPF prediction & Graph/tabular ICL $+$ LoRA & none (LLM predicts) & no \\
DrAgent~\cite{saha2025dragent} & Fault diagnosis & ReWOO planning & DR parsers & implicit \\
RePower~\cite{liu2025repower} & Research orchestration & Orchestrator $+$ retrieval & MATPOWER & implicit \\
\midrule
\textbf{This paper} & Forecasting, EV, power flow, contingency & Prompting $+$ agentic (solver-grounded) & CVXPY, PandaPower & \textbf{explicit gate $+$ safe failure} \\
\bottomrule
\end{tabularx}
\end{table}

\section{Validation Against a Published Agent}
\label{sec:appendix-validation}
As an external check, Table~\ref{tab:agent_comparison} compares GridDebugAgent (Section~\ref{sec:griddebug}) against the published X-GridAgent~\cite{wen2025xgridagent} on identical natural-language queries over standard IEEE systems. The two agents return the same bus connections, shortest-path sequences, and total active-power loss, confirming that the solver-grounded design reproduces established analysis results, even though our contingency-repair benchmark targets a different setting.

\begin{table}[t]
\centering
\caption{Comparison between Our Agent and X-GridAgent Outputs}
\label{tab:agent_comparison}
\scriptsize
\setlength{\tabcolsep}{3pt}
\begin{tabular}{|p{3.5cm}|p{5.9cm}|p{5.9cm}|}
\hline
\textbf{User Query} & \textbf{Our Agent Output} & \textbf{X-GridAgent Output} \\
\hline

Load the IEEE-39 bus system. What are the connected buses for line 11? What are the voltage limits of bus 3? &
For the IEEE-39 bus system:
Connected Buses for Line 11: Line 11 connects Bus 5 to Bus 10.
Voltage Limits of Bus 3: The voltage limits for Bus 3 are 0.94 pu (minimum) and 1.05 pu (maximum). &
- Connected buses for line index 11: from bus=5, to bus=10
- Voltage limits for bus 3: vm min=0.94 pu, vm max=1.05 pu. \\
\hline

For the IEEE 39-bus system, what is the shortest path between bus 1 and bus 10? &
From this connectivity, the shortest path from bus 1 to bus 10 is: Bus 1 $\rightarrow$ 2 $\rightarrow$ 3 $\rightarrow$ 4 $\rightarrow$ 5 $\rightarrow$ 10. &
- path bus indices: [1, 2, 3, 4, 5, 10]\newline
- path edges: [[1, 2], [2, 3], [3, 4], [4, 5], [5, 10]] \\
\hline

Run AC power flow on the
IEEE 39-bus system using the fast-decoupled method. Show
me the convergence result, the total active power loss, and
the active power flow value of line 11.&
Total Active Power Loss:\newline
The total active power generation is 6297.87 MW.
The total active power load is 6254.23 MW.
Therefore, the total active power loss is approximately ( 6297.87 - 6254.23 = 43.64 ) MW.
Active Power Flow on Line 11:\newline
Line 11, which runs from bus 5 to bus 10, has a loading of 67.2\%. This indicates the line is carrying a significant portion of its capacity, but the exact active power flow in MW would require further detailed data on line parameters and power factor. &
- Converged: ‘True’
- Total active power loss: 43.6423 MW
- Line index 11 active power flow: line 11 p from mw:
-322.6527 MW, line 11 p to mw: 323.3765 MW \\
\hline
\end{tabular}
\end{table}

\section{Design Checklist for Solver-Grounded Smart-Grid Agents}
\label{sec:appendix-checklist}
We distill the tutorial into a checklist a practitioner can apply when building an LLM or agent for a smart-grid task.
\begin{tcolorbox}[colback=gray!3,colframe=black,boxrule=0.5pt,sharp corners,breakable,left=2mm,right=2mm,top=1mm,bottom=1mm]
\footnotesize
\begin{enumerate}
\item \textbf{Separate the roles.} Let the LLM parse, retrieve, plan, orchestrate, and explain, never compute physical quantities. Assign every numerical result to a trusted solver or simulator (Section~\ref{sec:solver-grounded}).
\item \textbf{Define typed tool interfaces.} Fix inputs, units, output schemas, required parameters, and failure messages so the agent cannot silently mis-call a tool (Section~\ref{sec:action}).
\item \textbf{Specify the verification gate.} Enumerate the constraint-satisfaction checks (voltage/thermal limits, power balance, solver status) and the faithfulness checks (every reported number traces to tool output) that the final response must pass~\eqref{eq:verification}.
\item \textbf{Implement safe failure.} When verification fails, abstain and report the failure mode and its cause rather than fabricating, extrapolating, or silently repairing a value.
\item \textbf{Choose the orchestration unit by task structure.} A single-agent SDK for assistants, a multi-agent framework for genuinely distinct roles, or a workflow graph when stages must be verified before execution (Section~\ref{sec:frameworks}).
\item \textbf{Manage memory deliberately.} Keep a short-term scratchpad within a task and promote only verified results to long-term memory (Section~\ref{sec:memory}).
\item \textbf{Evaluate on all four metric groups.} Report task utility, solver-grounded correctness, faithfulness and safe failure, and cost and latency against an LLM-only baseline on identical data (Section~\ref{sec:evaluation}).
\item \textbf{Handle the no-solver case explicitly.} When no trusted solver exists, move verification to output-level checks (schema validity, value clipping) and expect prompt design to carry most of the operational weight.
\end{enumerate}
\end{tcolorbox}

\section{Additional Case-Study Visualizations}

\begin{figure}[htbp]
\centering
\includegraphics[width=0.9\linewidth]{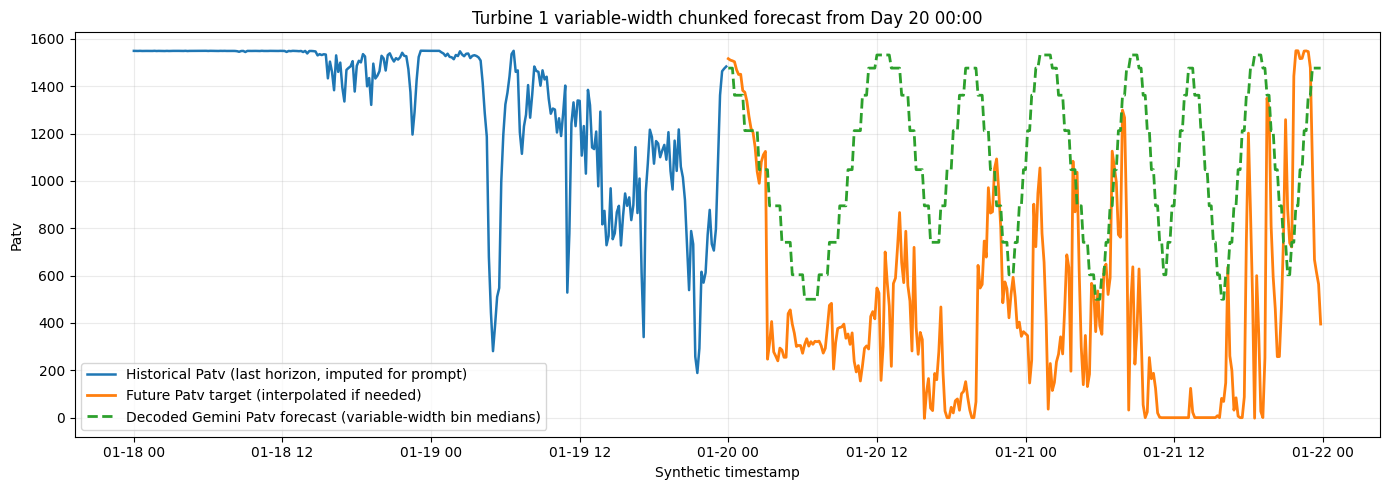}
\caption{Representative 48-hour wind power forecast for Turbine~1 on the SDWPF dataset (case study Section~\ref{sec:casestudies}-A). The solid blue curve is the historical active power ($P_{atv}$) used to build the prompt, the solid orange curve is the ground-truth future active power, and the dashed green curve is the LLM (Gemini) forecast decoded from variable-width bin medians. The LLM reproduces the broad diurnal pattern but overestimates output during low-generation periods and misses the sharp drops in the ground truth, illustrating the limits of LLM-only forecasting without a trusted solver.}
\label{fig:wind_forecast_ts}
\end{figure}

\begin{figure}[!htbp]
\centering
\includegraphics[width=0.75\linewidth]{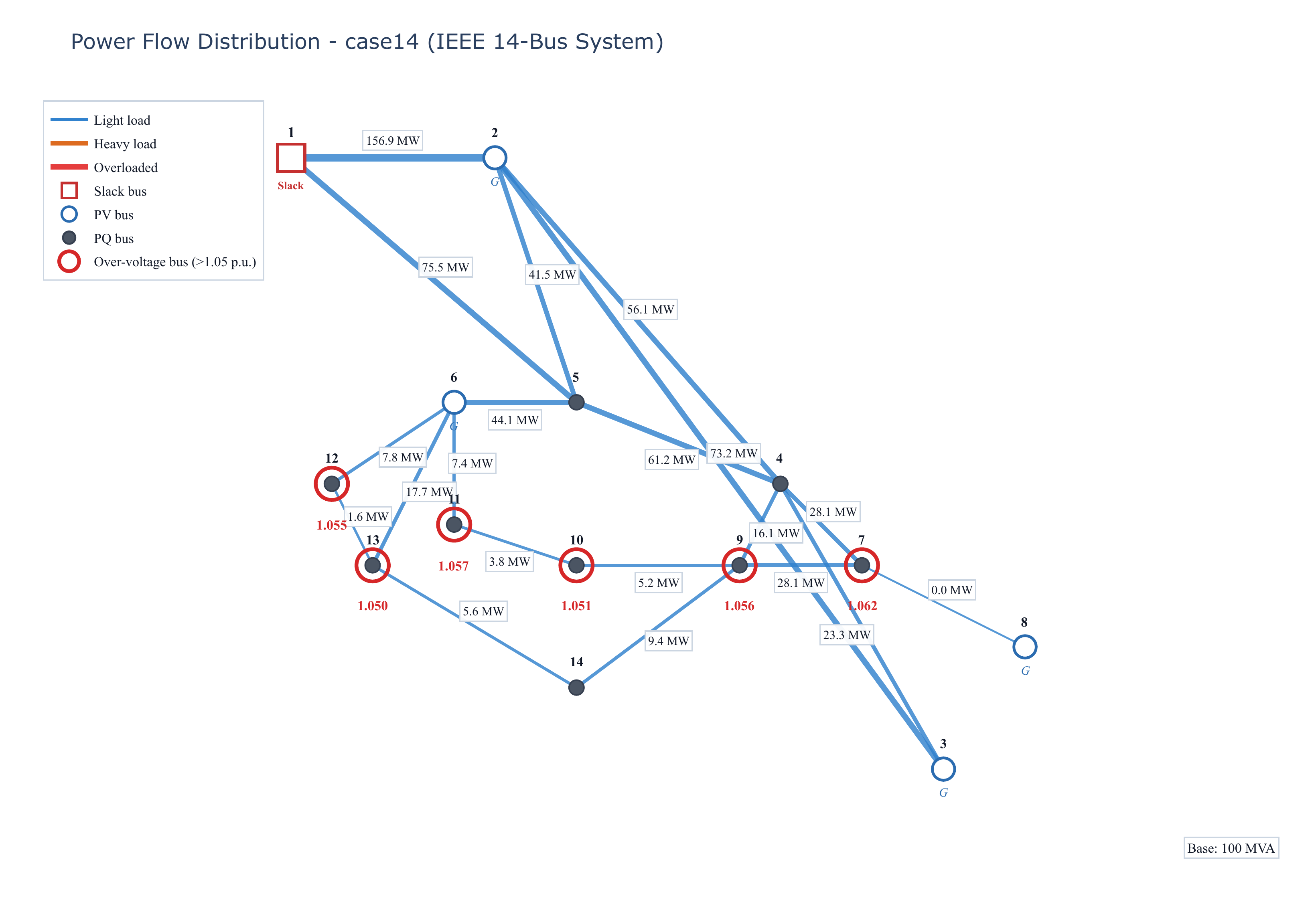}
\caption{IEEE-14 network with the PandaPower power-flow solution overlaid (case study Section~\ref{sec:pfagent}). Node markers distinguish the slack, PV, and PQ buses, and each branch is annotated with its active-power flow from the solver. The load buses whose solved voltage exceeds the 1.05 p.u. limit are ringed in red with their voltage magnitude, so the over-voltage conditions that the verification gate flags are directly identifiable on the topology. All values are the PandaPower solution, with a maximum branch flow of 156.9\,MW.}
\label{fig:ieee14_topology}
\end{figure}

\bibliographystyle{elsarticle-num}
\bibliography{refs}